\documentclass[10pt,journal,compsoc]{IEEEtran}

\usepackage{amsmath}
\usepackage{amssymb}
\usepackage[dvipdfmx]{graphicx}
\usepackage{color}
\usepackage{dsfont}

\usepackage{supertabular}
\usepackage{comment}

\ifCLASSOPTIONcompsoc

\usepackage{cite}
\usepackage{url}

\usepackage{theorem}
\theorembodyfont{\rmfamily}

\newtheorem{theorem}{Theorem}
\newtheorem{lemma}{Lemma}

\newcommand{\bm}[1]{\boldsymbol{#1}}

\DeclareMathOperator{\sfdiag}{\mathsf{diag}}
\DeclareMathOperator{\sfcol}{\mathsf{col}}

\DeclareMathOperator{\sftrace}{\mathsf{trace}}

\DeclareMathOperator{\sfker}{\mathsf{ker}}
\DeclareMathOperator{\sfim}{\mathsf{im}}

\DeclareMathOperator{\sfvec}{\mathsf{vec}}

\newcommand{\mat}[1]{\left[\: \begin{matrix} #1 \end{matrix} \:\right]}

\newcommand{\spliteq}[1]{\begin{split} #1 \end{split}}
\newcommand{\simode}[1]{\left\{\:  \begin{aligned} #1 \end{aligned} \right.}

\begin{document}

\title{
Explicit Ensemble Mean Clock Synchronization for Optimal Atomic Time Scale Generation
}

\author{Takayuki Ishizaki$^1$, 
Takahiro Kawaguchi$^2$, Yuichiro Yano$^3$, Yuko Hanado$^3$ 
\thanks{$^1$ Institute of Science Tokyo, 2-12-1, Ookayama, Meguro, Tokyo, 152-8552, Japan}
\thanks{$^2$ Gunma University, 1-5-1, Tenjincho, Kiryu, Gunma, 376-8515, Japan}
\thanks{$^3$ National Institute of Information and Communications Technology, 4-2-1, Nukui-Kitamachi, Koganei, Tokyo, 184-8795 , Japan}
\thanks{Manuscript received April 19, 2005; revised August 26, 2015.}
}

\IEEEtitleabstractindextext{%
\begin{abstract}
This paper presents a novel theoretical framework, called explicit ensemble mean (EEM) synchronization.
This framework unifies time scale generation, clock synchronization, and oscillator frequency regulation within the systems and control theory paradigm.
By exploiting the observable canonical decomposition of a standard atomic ensemble clock model, the system is decomposed into two complementary components: the observable part, which represents the synchronization error, and the unobservable part, which captures the synchronization destination. 
Within this structure, we mathematically prove that standard Kalman filtering, which is widely used in current time scale generation, not only performs observable state estimation, but also significant unobservable state estimation,  and it can be interpreted as a special case of the proposed framework that optimizes long-term frequency stability in terms of the Allan variance. 
Furthermore, applying state feedback control based on Kalman filtering to each component achieves optimal time scale generation, clock synchronization, and oscillator frequency regulation in a unified manner.
The proposed framework provides a foundation for developing explainable timing systems.
\end{abstract}

}

\maketitle

\section{Introduction}

\subsection{Background}

Time scale generation from an ensemble of atomic clocks has long been a cornerstone technique in time and frequency metrology, particularly in the construction of national and international time references such as Coordinated Universal Time (UTC)\cite{galleani2020generating,nemitz2021absolute,lee2023characteristics}. 
In addition, this approach is critical to realizing resilient timing infrastructures that can operate robustly even without reference time derived from Global Navigation Satellite System (GNSS) signals. 
Examples include composite clocks and the continuously operating primary reference time clock, the Coherent Network Primary Reference Time Clock (cn-PRTC)\cite{galleani2003use,cosart2022cnprtc}.

In practice, a time scale, or standard time reference, called the atomic time (TA), is generated and maintained by 
physically regulating the frequency of local oscillators and digitally adjusting clock readings
based on the ensemble time generated from multiple pairwise clock phase difference measurements \cite{arias2011timescales}. 
While a number of studies have examined methods for computing TA as a reference, few have addressed integrated system design, including the regulation and steering of physical oscillators. 
In particular, a unified theoretical framework based on systems and control theory that systematically addresses time scale generation, clock synchronization, and oscillator frequency regulation yet to be established. 
Furthermore, little attention has been given to mathematically proving the working principles and convergence properties of such an integrated system. 
These issues are critical to the stability of reliable timing infrastructures in the short and long term.

Kalman filtering is widely used in this context, such as in the GPS composite clock \cite{brown1991theory},
due to its ability to provide optimal state estimation under Gaussian noise assumptions \cite{galleani2010time}. 
The Kalman filtering has been used to generate a time scale by implicitly computing the ensemble mean weight of the clock states, known as the ``implicit" ensemble mean. 
However, the direct application of Kalman filtering has certain limitations. 
In particular, the error covariance matrix diverges over time, even though the Kalman gain converges to a constant value. 
This divergence implies the presence of indeterminate operations in the algorithm, which can lead to numerical instability.
Several error covariance reduction schemes have been proposed to address this issue \cite{brown1991theory,greenhall2012review}, but it has been observed that optimal Kalman filtering often results in suboptimal time scales in terms of short-term frequency stability.
An ad hoc method involving the modification of the Kalman gain has been introduced to mitigate this problem \cite{greenhall2006kalman}, called the Kalman Plus Weights (KPW) algorithm, though the underlying mechanism has not been theoretically elucidated.

In addition to approaches that implicitly determine the ensemble mean weight of atomic clocks through Kalman filtering, there are also approaches that use Kalman filtering solely to estimate the clock state or noise. 
These approaches assign the ensemble mean weight according to a different policy.
One example is the KAS-2 algorithm \cite{stein2003time,erickson2011symmetricom}. 
This method avoids error covariance divergence by applying the Kalman filtering only to estimate relative or  differential clock states. 
The ensemble mean weight of the clocks is indirectly determined by the technical assumption that the weighted average of the stochastic clock noise is always zero.
Another example is the AT1 algorithm and its extended version, the AT2 algorithm \cite{weiss1991at2,levine2012invited}.
These methods update the ensemble mean weights based on the magnitude of the state estimation error during operation. 
The weights are known to be updated in a manner that improves short-term stability. 
Kalman filtering is used for clock frequency estimation in the AT2 algorithm.
Note that these time scale generation algorithms use Kalman filtering to digitally adjust clock readings rather than physically regulating the frequency of local oscillators. 
In other words, they perform so-called paper clock calculations.

\subsection{Contributions}

In this paper, we present a unified and rigorous framework called ``explicit" ensemble mean (EEM) synchronization.
This framework integrates time scale generation, clock synchronization, and oscillator frequency regulation within a coherent systems and control theory paradigm. 
This framework provides an explicit understanding of how the Kalman filtering for the GPS composite clock
implicitly computes the ensemble mean weight of clocks.
In particular, we prove that the Kalman filtering is significant not for estimating the observable relative clock state, but rather for estimating the ``unobservable" weighted average state.
From a systems and control theory perspective, it is counterintuitive that the Kalman filtering is significant for estimating unobservable states.
The literature lacks such a clear explanation of the working principles of implicit ensemble mean for time scale generation.

The main contributions are summarized as follows.

\begin{itemize}
    \item We derive an equivalent formulation of the Kalman filtering algorithm that completely eliminates indeterminate operations leading to the error covariance divergence by applying a specially tailored observable canonical decomposition to a standard atomic ensemble clock model.
    \item We construct a physical oscillator frequency regulation algorithm that  synchronizes all clocks in the ensemble to a free-running ensemble mean dynamics, the weighting of which can be explicitly specified by the user.
    \item We mathematically prove that the standard Kalman filtering
    is a special case of the proposed EEM synchronization, which achieves the least Allan variance only in the long term.
    \item We design a novel synchronization algorithm that explicitly balances both short-term and long-term frequency stability of the generated time scale by incorporating intermittent feedback control of the synchronization destination.
\end{itemize}
These contributions provide a rigorous foundation for resilient timing infrastructure and open new avenues for atomic time scale generation, clock synchronization, and oscillator frequency regulation via advanced systems and control theory.

\subsection{Mathematical Preliminaries}

\subsubsection{Symbols}


Throughout the paper, we use the following symbols.
\[
\begin{array}{ll}
\mathbb{R} & \mbox{the set of real numbers} \\
\mathbb{Z} & \mbox{the set of integers } \\
\mathds{1}_N & \mbox{the $N$-dimensional all-ones vector} \\
I_N & \mbox{the $N$-dimensional identity matrix} \\
A^{\sf T} & \mbox{the transpose of a matrix $A$} \\
\end{array}
\]
\[
\begin{array}{ll}
\sfim A & \mbox{the image of a matrix $A$} \\
\sfker A & \mbox{the kernel of a matrix $A$} \\
\bm{\Lambda}(A) & \mbox{the set of the eigenvalues of a square matrix $A$} \\
\mathbb{E}[X] & \mbox{the mean of a random variable $X$} 
\end{array}
\]
Other symbols will be explained when used.
We use the equality  ``$:=$" in the sense that a new symbol on the left is defined by known symbols on the right.

The Kronecker product of matrices is denoted by
\[
A \otimes B := \mat{
a_{11}B & \cdots & a_{1n}B \\
\vdots & \ddots & \vdots \\
a_{m1}B & \cdots & a_{mn} B
}
\]
where $a_{ij}$ is the $(i,j)$-element of $A$.
In this paper, we use the property of multiplication
\[
(A \otimes B) (C \otimes D) = AC \otimes BD.
\]
For a sequence $\{x[k]\}_{k\in \mathbb{Z}}$, the first-order difference operation is denoted by
\[
\triangle_m x[k] := x[k+m] - x[k]
\]
where $m$ denotes an interval.
In the same way, the second-order difference operation is denoted by
\[
\triangle^2_m x[k] := x[k+2m] - 2x[k+m] + x[k],
\]
which is equal to $\triangle_m(\triangle_m x[k])$.

\subsubsection{Gaussian Processes}

A sequence $\{x[k]\}_{k\in \mathbb{Z}}$ of random variables 
is said to be a Gaussian process if the whole sequence follows a multivariate Gaussian distribution.
Furthermore, a Gaussian process is said to be stationary if there exist some constant parameters $\mu$ and $r_{ij}$ for $(i,j) \in \mathbb{Z}^2$ such that
\[
\mathbb{E}\bigl[x[k] \bigr] = \mu
,\quad
\mathbb{E}\bigl[ (x[k+i]-\mu) (x[k+j]-\mu)^{\sf T} \bigr] = r_{ij}
\]
for any $k \in \mathbb{Z}$.
In particular, it is said to be white if the mean and time correlation are zero, namely
\[
\mu =0 ,\quad
r_{ij} = 0,\quad 
\forall i \neq j .
\]
In this paper, the family of zero-mean stationary Gaussian processes is denoted by $\mathcal{G}$.

Consider a discrete-time linear system 
\[
x[k+1] = A x[k] + v[k]
\]
driven by a white Gaussian process $\{v[k]\}_{k\in \mathbb{Z}}$, where $A$ is a square matrix and $x[k]$ is a state.
Clearly, the state sequence $\{x[k]\}_{k\in \mathbb{Z}}$ is a Gaussian process.
In addition, the zero-mean stationarity of the state process is equivalent to system stability in the sense that
\[
\{x[k]\}_{k\in \mathbb{Z}}  \in \mathcal{G}
\quad \iff \quad
\bm{\Lambda}(A )\subseteq \mathds{D}
\]
where the unit disk in the complex plane is denoted by
\[
\mathds{D}:= \{z \in \mathbb{C}: |z| < 1\}.
\]
In such a way, the symbol $\mathcal{G}$ will be used to represent the system stability, or the boundedness of the variance.


\section{Problem Setting}

\subsection{Single Atomic Clock Model}

\subsubsection{Continuous-Time Model}

In this paper, we discuss a standard second-order atomic clock model \cite{galleani2008tutorial,galleani2010time}, although similar analysis can be applied also to higher-order clock models.
It is known from experimental observations that the time sequence of clock reading deviation relative to an ideal time coordinate can be  modeled as a multiple integration of white Gaussian processes.
In particular, the second-order model is given as 
\begin{equation}\label{eq:intmodel}
h(t) = c_1 + c_2 t 
     + \int_0^t v_1(t_1) dt_1 + \int_0^t \int_0^{t_1} v_2(t_2) d t_2 dt_1
\end{equation}
where $c_i$ is a scalar constant of the initial value, and  
\[
v_i=\{v_i(t): t \in \mathbb{R} \}
\]
denotes a continuous-time white Gaussian process such that
\begin{equation}
\spliteq{
    & \mathbb{E}\bigl[ v_i(t) \bigr] =0 \\
    & \mathbb{E}\bigl[ v_i(t)v_i(t+\tau) \bigr] = \sigma_i^2 \delta(\tau) \\
    & \mathbb{E}\bigl[ v_1(t)v_2(t+\tau) \bigr] = 0
    ,\quad \forall \tau \in \mathbb{R},
    }
\end{equation}
where $\delta(\cdot)$ denotes the Dirac delta function. 
The integral equation model in \eqref{eq:intmodel} can be represented by a system of stochastic differential equations as
\begin{equation}\label{eq:diffeqs}
\simode{
	\dot{x}_1(t) & = x_2(t) + v_1(t) \\
	\dot{x}_2(t) & = v_2(t) + u(t),
}
\end{equation}
where a control input $u(t)$ is introduced to discuss the frequency regulation of physical oscillators.
If the initial values are given such that
\[
x_{1}(0) = c_{1}
,\quad x_{2}(0) = c_{2},
\]
and the control input is not applied, then the first state variable $x_1(t)$ represents the clock reading deviation of the free-running clock as
\[
x_1(t) = h(t),\quad \forall t \in \mathbb{R}.
\]

Furthermore, consider the stacked vectors denoted by
\[
x(t):=\mat{x_1(t) \\ x_2(t)}
,\quad
v(t):=\mat{v_1(t) \\ v_2(t)},
\]
where the first elements are related to phase and the second elements to frequency.
Then, we can write the stochastic differential equation system in \eqref{eq:diffeqs} in the matrix form as
\begin{equation}\label{eq:ssmodel}
\simode{
	\dot{x}(t) &= A_{\rm c} x(t) + B_{\rm c}u(t) + v(t) \\
	h(t) &= C_{\rm c} x(t)
}
\end{equation}
where $A_{\rm c}\in \mathbb{R}^{2\times 2}$, $B_{\rm c}\in \mathbb{R}^{2}$, and $C_{\rm c}\in \mathbb{R}^{1\times 2}$ are defined as
\begin{equation}\label{eq:ssmat}
A_{\rm c}:=
\mat{
    0 & 1\\
    0 & 0
}
,\quad
B_{\rm c}:= \mat{0\\1}
,\quad
C_{\rm c} := \mat{1 & 0}.
\end{equation}
The subscript ``${\rm c}$" of the system matrices indicates that the model is continuous-time.
The process noise 
\[
v=\{v(t): t \in \mathbb{R} \}
\]
is a white Gaussian process with a constant covariance such that
\[
\mathbb{E}\bigl[ v(t) \bigr] =0
,\quad
\mathbb{E}\bigl[ v(t) v^{\sf T}(t+\tau) \bigr] = 
Q_{\rm c} \delta(\tau)
,\quad \forall t \in \mathbb{R}.
\]
where the covariance matrix is diagonal and given as 
\[
Q_{\rm c}:= \mat{
\sigma_1^2 & 0 \\
0 & \sigma_2^2
}.
\]
Note that $\sigma_1$ is the standard deviation of the white frequency noise, and $\sigma_2$ is that of the random walk frequency noise.
This is known as a cesium-type atomic clock model.
Due to the integration of white frequency noise and random walk frequency noise, the variance of the clock reading deviation $h(t)$ diverges over time.

\subsubsection{Equivalent Discrete-Time Model}

Consider a discrete-time sequence in an ideal time coordinate.
The interval between two adjacent times is assumed to be constant, or equivalently 
\begin{equation}\label{eq:distime}
t_k := \tau k
,\quad k \in \mathbb{Z}
\end{equation}
where $\tau$ denotes a sampling interval.
The relation between $x(t_k)$  and $x(t_{k+1})$ in \eqref{eq:ssmodel}
can be formally expressed as 
\[
x(t_{k+1}) = 
e^{A_{\rm c} \tau} x(t_k) + \!
\int_{t_k}^{t_{k+1}} \!\!
e^{A_{\rm c}(t_{k+1}-t)}\bigl(B_{\rm c} u(t) + v(t)\bigr)dt.
\]
Suppose that a zero-order hold input is applied, namely
\[
u(t)=
u[k], \quad 
\forall t \in [k\tau, (k+1)\tau).
\]
Then, we can derive a discrete-time model that is equivalent to the continuous-time model in \eqref{eq:ssmodel} over the discrete time sequence as
\begin{equation}\label{eq:ssmodeld}
    \simode{
    x[k+1] &= A x[k] + Bu[k] + v[k] \\
    h[k] &= C x[k]}
\end{equation}
where $A\in \mathbb{R}^{2\times 2}$, $B\in \mathbb{R}^{2}$, and $C\in \mathbb{R}^{1\times 2}$ are defined as
\begin{equation}\label{eq:Atau}
A := \mat{
1 & \tau \\
0 & 1
}
,\quad
B := \mat{
\tau \\
1
}
,\quad
C:= \mat{1 & 0},
\end{equation}
the discrete-time state and clock reading deviation as
\[
x[k]:=x(t_k)
,\quad
h[k]:= h (t_k).
\]
Furthermore, the process noise $\{v[k]\}_{k\in \mathbb{Z}}$
is a white Gaussian process such that 
\[
\mathbb{E}\bigl[ v[k] \bigr] =0
,\quad
\mathbb{E}\bigl[ v[k] v^{\sf T}[k] \bigr] = Q,
\quad 
\forall k \in \mathbb{Z}
\]
where the covariance matrix is given as
\begin{equation}\label{eq:defQk}
Q :=
\mat{
\tau \sigma_1^2 + \frac{\tau^3}{3} \sigma_2^2   & \frac{\tau^2}{2}\sigma_2^2  \\
\frac{\tau^2}{2} \sigma_2^2 & \tau \sigma_2^2 
}.
\end{equation}

The clock model in \eqref{eq:ssmodeld} involves the physical regulation of oscillator frequency.
In contrast, the digital adjustment of the clock reading deviation can be expressed as 
\begin{equation}\label{eq:digmod}
\simode{
x'[k+1] &= A x'[k] + v[k] \\
h'[k] &= C x'[k] + u'[k]
}
\end{equation}
where $x'[k]$ is the free-running clock state, $u'[k]$ is the clock reading adjustment, and $h'[k]$ is the adjusted clock reading deviation.
Note that while any physical oscillator frequency regulation can be imitated by a digital adjustment to the clock reading, the reverse is not true.
This can be seen as follows.
For any given sequence $\{u[k]\}_{k\in \mathbb{Z}}$, we can match $\{h'[k]\}_{k\in \mathbb{Z}}$ in \eqref{eq:digmod} to $\{h[k]\}_{k\in \mathbb{Z}}$ in \eqref{eq:ssmodeld}  if $\{u'[k]\}_{k\in \mathbb{Z}}$ is constructed by
\[
\simode{
\epsilon[k+1] &= A \epsilon[k] + B u[k] \\
u'[k] &= C \epsilon[k] 
}
\]
where $\epsilon[k]$ is the digital imitation state such that
\[
x[k]=x'[k] + \epsilon[k].
\]
However, for a given sequence $\{u'[k]\}_{k\in \mathbb{Z}}$, it does not necessarily exist $\{u[k]\}_{k\in \mathbb{Z}}$ such that $\{h[k]\}_{k\in \mathbb{Z}}$ is matched to $\{h'[k]\}_{k\in \mathbb{Z}}$.
This is simply because the frequency regulation in \eqref{eq:ssmodeld} can affect the clock reading via clock dynamics.
In this sense, the physical regulation of oscillator frequency, as discussed in this paper, is generally more difficult than the digital adjustment of clock readings, i.e., paper clock calculations.
The latter is considered in most time scale generation algorithms, such as the GPS composite clock, KPW , KAS-2, and AT1/AT2 \cite{brown1991theory,greenhall2006kalman,galleani2010time,greenhall2012review,stein2003time,erickson2011symmetricom,weiss1991at2,levine2012invited}.

\subsection{Ensemble Atomic Clock Model}

Consider an ensemble of $N$ clocks, each of which is written as the single clock model in \eqref{eq:ssmodeld}.
In the following, the state and process noise of the $i$th clock are denoted by
\[
x_i[k] :=\mat{
x_{1i}[k] \\
x_{2i}[k] 
}
,\quad
v_i[k] :=\mat{
v_{1i}[k] \\
v_{2i}[k] 
}
\]
where the first elements are relevant to the phase and the second elements to the frequency.
With a slight abuse of notation, we use the same symbols as those for the single clock model to represent the stacked versions
\[
x[k]:= \mat{
\sfcol (x_{1i}[k])_{i\in \mathds{N}} \\
\sfcol (x_{2i}[k])_{i\in \mathds{N}} 
}
,\quad
v[k]:= \mat{
\sfcol (v_{1i}[k])_{i\in \mathds{N}} \\
\sfcol (v_{2i}[k])_{i\in \mathds{N}} 
} 
\]
where $\mathds{N}:=\{1,\ldots,N\}$ denotes the label set of clocks, and $\sfcol(\cdot)$ denotes the stacked column vector.
In the same way, we denote the control input to the $i$th clock by $u_i[k]$, the clock reading deviation of the $i$th clock by $h_i[k]$, and 
\[
u[k] :=
\sfcol (u_{i}[k])_{i\in \mathds{N}}
,\quad
h[k] :=
\sfcol (h_{i}[k])_{i\in \mathds{N}}.
\]
Then, we have the model of the ensemble atomic clock as
\begin{equation}\label{eq:Csensm}
\simode{
x[k+1]&= (A \otimes I_N) x[k] + (B \otimes I_N) u[k] + v[k] \\
y[k] &= (C \otimes V) x[k] + w[k] \\
h[k] &= (C \otimes I_N) x[k]
}
\end{equation}
where $y[k]$ is the measurement of the relative clock readings among the clocks,
and
$V \in \mathbb{R}^{(N-1)\times N}$ denotes a matrix that represents pairs of clocks whose phase difference is being measured.
For example, the choice of
\begin{equation}\label{eq:mesV}
V = 
\mat{
I_{N-1} & -\mathds{1}_{N-1}
}
\end{equation}
corresponds to the case where the phase differences between the $N$th clock and others are measured 
in such a way that
\[
y_i[k] = h_i[k] -h_N[k] + w_i[k] , \quad i \in \mathds{N}^-
\]
where the label set without $N$ is denoted by
\[
\mathds{N}^-:=\{1,\ldots,N-1\}.
\]
Note that the formulation of $V$ in \eqref{eq:Csensm} allows any set of $N-1$ clock pairs for measuring phase differences or relative clock readings.
Furthermore, the measurement noise $\{w[k]\}_{k\in \mathbb{Z}}$ is supposed to be a white Gaussian process such that
\[
\mathbb{E}\bigl[ w[k] \bigr] =0
,\quad
\mathbb{E}\bigl[ w[k] w^{\sf T}[k] \bigr] = \bm{R},
\quad 
\forall k \in \mathbb{Z},
\]
where $\bm{R}\in \mathbb{R}^{(N-1)\times (N-1)}$ is a positive definite covariance matrix.
Supposing that each atomic clock is independent of the others, we have the covariance matrix $\bm{Q}\in \mathbb{R}^{2N\times 2N}$ of the process noise as
\begin{equation}\label{eq:CscovQ}
\bm{Q} := \mat{
\tau \Sigma_1 + \frac{\tau^3}{3} \Sigma_2   & \frac{\tau^2}{2}\Sigma_2  \\
\frac{\tau^2}{2} \Sigma_2 & \tau \Sigma_2 
}
\end{equation}
where $\Sigma_1 \in \mathbb{R}^{N \times N}$ and $\Sigma_2 \in \mathbb{R}^{N \times N}$ are defined as
\[
\Sigma_1 := 
\sfdiag(\sigma_{11}^2,\ldots,\sigma_{1N}^2)
,\quad
\Sigma_2 := 
\sfdiag(\sigma_{21}^2,\ldots,\sigma_{2N}^2).
\]
Note that $\sigma_{1i}$ and $\sigma_{2i}$ denote the standard deviations of the white frequency noise and random walk frequency noise of the $i$th clock, respectively.

The ensemble clock model in \eqref{eq:Csensm} has two outputs $y[k]$ and $h[k]$.
The former is the noisy measurement data used in a time scale generation algorithm.
The latter is the theoretical value of the clock reading deviations used in the theoretical evaluation of a generated time scale.
We will use these two outputs differently based on this distinction.

\subsection{Kalman Filtering for Time Scale Generation}

\subsubsection{Kalman Filtering Algorithm}

We consider state estimation of the ensemble clock model in \eqref{eq:Csensm} based on the Kalman filtering \cite{kalman1960new}.
To simplify the notation, we denote the ensemble system matrices by
\[
\bm{A}:=  A \otimes I_N,
\quad
\bm{B}:=  B \otimes I_N,
\quad
\bm{C}:=  C \otimes V.
\]
In the following, the prior and posterior state estimates are, respectively, denoted by 
\[
\hat{x}^-[k] := \mathbb{E} \bigl[ x[k]  | Y[k-1] \bigr] ,\quad
\hat{x}[k] := \mathbb{E} \bigl[ x[k]  | Y[k] \bigr], 
\]
where we consider the expectation of the state conditional on the past measurement information
\[
Y[k]:=\{y[k],y[k-1],\ldots\}.
\]
The corresponding error covariances are defined as
\[
\spliteq{
\bm{P}^- [k] &:= \mathbb{E} \bigl[ (x[k] - \hat{x}^-[k]) (x[k] - \hat{x}^-[k])^{\sf T}  \bigr], \\
\bm{P}[k] &:= \mathbb{E} \bigl[ (x[k] - \hat{x}[k]) (x[k] - \hat{x}[k])^{\sf T} \bigr].
}
\]
Then, the Kalman filtering algorithm is obtained as
\begin{equation}\label{eq:kalfil}
\simode{
\hat{x}^-[k] &= \bm{A} \hat{x}[k-1] + \bm{B} u[k-1] \\
\bm{P}^- [k] &= \bm{A} P[k-1] \bm{A}^{\sf T} + \bm{Q}\\
\bm{H}[k] &= \bm{P}^-[k] \bm{C}^{\sf T} \bigl( \bm{C} \bm{P}^-[k] \bm{C}^{\sf T} + \bm{R} \bigr)^{-1} \\
\bm{P}[k] &= \bigl( I_{2N} - \bm{H}[k] \bm{C} \bigr) \bm{P}^-[k] \\
\hat{x}[k] &= \hat{x}^-[k] + \bm{H} [k] \bigl(y[k] - \bm{C}\hat{x}^-[k] \bigr) ,
}
\end{equation}
where $\bm{H}[k]$ is called the Kalman gain.
This algorithm minimizes the variance of the state estimation error
\[
\mathbb{E} \left[ \bigl\| \bm{G} e[k] \bigr\|^2 \right]
= \sftrace \bigl( \bm{G} \bm{P}[k] \bm{G}^{\sf T} \bigr)
\]
for any matrix $\bm{G}$, where the state estimation error is 
\[
e[k]:=x[k] - \hat{x}[k].
\]
This is known as the optimality of conditional expectation.

The Kalman filtering algorithm in \eqref{eq:kalfil} is used as the time scale generation algorithm for the GPS composite clock \cite{brown1991theory}.
Unless otherwise stated, this paper will focus on this Kalman filtering algorithm for time scale generation.

\subsubsection{Application to Time Scale Generation}

For an observable system, the Kalman filtering algorithm produces the state estimate such that
\[
\{ e[k] \}_{k \in \mathbb{Z}} \in \mathcal{G} 
\]
in steady state, or more specifically
\[
\lim_{k\rightarrow \infty} \mathbb{E} \left[ e[k] \right] =0
,\quad
\lim_{k\rightarrow \infty} \bm{P}[k] = \bm{P}^{\star},
\]
where $\bm{P}^{\star}$ is a positive semidefinite matrix.
However, this does not hold for the ensemble clock model in \eqref{eq:Csensm}.
This is because the ensemble clock model is unobservable; the ``relative" clock readings $y[k]$ among the clocks can only be used to estimate the ensemble clock state $x[k]$.
In particular, there is a nontrivial unobservable subspace
\begin{equation}\label{eq:unobsss}
\bigcup_{\lambda \in \mathbb{C}}
\sfker 
\mat{
\lambda I_{2N} - \bm{A} \\
\bm{C}
}
=
\sfim  (I_2 \otimes \mathds{1}_N).
\end{equation}
This means that the subspace such that the phase and frequency of all clocks are synchronized, namely  
\[
x_{1}[k]=x_{2}[k]=\cdots=x_{N}[k]
\]
cannot be estimated through measurement of the relative clock readings.
In systems and control theory, the observability for linear systems can be defined as the condition that the initial state is uniquely identified by the measurement sequence, provided that the process noise and measurement noise are zero.
Several equivalent definitions and conditions of the observability are known for linear systems \cite{fairman1998linear}.

Due to the unobservability, the state estimation error of the ensemble clock model diverges over time along the unobservable subspace.
In other words, although the state estimation error diverges, its synchronization error disappears with time.
Mathematically, it can be represented as
\[
\{ (I_2 \otimes V) e[k] \}_{k \in \mathbb{Z}} \in \mathcal{G}.
\]
For example, the specific choice of $V$ in \eqref{eq:mesV} implies that the relative state estimation error $\{\Delta e_i^N[k]\}_{k\in \mathbb{Z}}$ is a zero-mean stationary Gaussian process for all $i\in \mathds{N}^-$ where
\[
\Delta e_i^N[k]:=e_{i}[k]-e_{N}[k].
\]
Because $V \mathds{1}_N$ must be zero from definition, we see that
\[
\sfker (I_2 \otimes V) = \sfim (I_2 \otimes \mathds{1}_N).
\]
This means that  multiplying $e[k]$ by $(I_2 \otimes V)$ eliminates the component of the synchronized estimation error associated with the unobservable subspace.
This does not depend on the particular choice of $V$.
Thus, stabilizing the ensemble state estimate $\hat{x}[k]$ by the control input $u[k]$ so that
\[
\{ \hat{x}[k] \}_{k \in \mathbb{Z}} \in \mathcal{G}
\quad \iff \quad
\{  x[k] - e[k] \}_{k \in \mathbb{Z}} \in \mathcal{G},
\]
we can realize the clock synchronization in the sense that
\[
\{ (I_2 \otimes V) x[k] \}_{k \in \mathbb{Z}} \in \mathcal{G}
\]
by the physical oscillator frequency regulation.
For example, the specific choice of $V$ in \eqref{eq:mesV} implies that the relative clock state $\{\Delta x_i^N[k]\}_{k\in \mathbb{Z}}$ is a zero-mean stationary Gaussian process for all $i\in \mathds{N}^-$ where
\begin{equation}\label{eq:relst}
\Delta x_i^N[k]:=x_{i}[k]-x_{N}[k].
\end{equation}
Note that the statistical properties of the synchronization destination, such as the magnitude of the variance and the rate of divergence, are  comparable to those of the state estimation error. 
This can be seen from the facts that
\begin{itemize}
\item by the stabilization of the ensemble state estimate $\hat{x}[k]$, the ensemble clock state $x[k]$ converges to the state estimation error $e[k]$ in the sense of zero mean and finite variance for their relative values, and
\item both the mean and variance of the state estimation error $e[k]$ diverge over time.
\end{itemize}
In this sense, analyzing the state estimation error is essential for time scale generation.

\begin{table*}[t]
\caption{Standard deviations of white frequency noise and random walk frequency noise.}
\label{table:noisepara}
\centering
\begin{tabular}{lccccccccccccccc}
& scale & 1  &  2 &   3  &  4  &  5  &  6  &  7  &  8  &  9  &  10  \\
\hline
$\sigma_1$ & $ 10^{-9}$ & 0.1700  &  0.0886 &   0.1221  &  0.1273  &  0.2185  &  0.1063  &  0.1805  &  0.2168  &  0.0930  &  0.1801  \\
$\sigma_2$ &  $ 10^{-12}$ &  0.1507  &  0.0532  &  0.0167  &  0.0771  &  0.2940  &  0.0492  &  0.0407  &  0.0829  &  0.0520  &  0.0566  \\  \hline
\end{tabular}
\end{table*}

\begin{table*}[t]
\caption{Standard deviations of measurement noise.}
\label{table:noiseparam}
\centering
\begin{tabular}{lccccccccccccccc}
scale & (1,10)  &  (2,10) &   (3,10)  &  (4,10)  &  (5,10)  &  (6,10)  &  (7,10)  &  (8,10)  &  (9,10)   \\
\hline
 $10^{-14}$ & 0.4353  &  0.0759  &  0.4720  &  0.1166  &  0.4148  &  0.0885  &  0.0998  &  0.2453  &  0.0373  \\
 \hline
\end{tabular}
\end{table*}

\subsection{Problems to Discuss}

\subsubsection{The Allan Variance}

Suppose that the control input $\{u[k]\}_{k\in \mathbb{Z}}$ is zero for a single clock model in \eqref{eq:ssmodeld}.
The Allan variance \cite{allan1966statistics} is defined by
\[
\sigma_{\sf A}^2(\tau) := \mathbb{E}\left[
\frac{ (\triangle^2_1 h[k])^2}{2\tau^2}
\right]
\]
where the clock model should be considered as an implicit function of the sampling interval $\tau$.
In fact, the Allan variance of the free-running clock model is a function only of $\tau$, and it can be analytically found \cite{zucca2005clock,ishizaki2024higher} as
\begin{equation}\label{eq:anAllan}
\sigma_{\sf A}^2(\tau) =
\frac{1}{\tau}\sigma_1^2 + \frac{\tau}{3} \sigma_2^2 ,
\end{equation}
where $\sigma_1$ is the standard deviation of the white frequency noise, and $\sigma_2$ is that of the random walk frequency noise.
In this paper, we call such an explicit representation the analytical Allan variance.

On the other hand, when the control input is applied, the Allan variance is generally not a function of $\tau$ alone, nor can it be expressed analytically.
In such a case, a statistical estimator of the Allan variance is computed from a measurement data set.
Suppose that an output sequence
\[
\{h[k]\}_{k\in \mathds{T}} := \{h[0],h[1],\ldots,h[T] \}
\]
is measured over the sampling interval $\tau$.
Then, a statistical estimator of the Allan variance can be computed as
\[
\hat{\sigma}_{\sf A}^2 (m; \{h[k]\}_{k\in \mathds{T}}) := 
\frac{1}{T-2m} \sum_{k=0}^{T-2m-1} 
\frac{
\left( \triangle_{m}^2 h[k]  \right)^2
}{
2 (m \tau)^2  
}
\]
where $m$ is an interval, and the set of feasible values is
\[
\mathds{M} := \left\{
1,2,\ldots, \left\lfloor \tfrac{T-1}{2} \right\rfloor
\right\}.
\]
In this paper, we call this estimate the statistical Allan variance.
Typically, the statistical Allan variance are presented as a double logarithmic plot of 
\[
\mathcal{L}\{h[k]\}_{k\in \mathds{T}}:=
\left(
\tau m, 
\hat{\sigma}_{\sf A}^2 (m; \{h[k]\}_{k\in \mathds{T}})
\right)_{m\in \mathds{M}}.
\]
This estimation is used as standard for the evaluation of generated time scales.
In the following, for a vector-valued sequence $\{h[k]\}_{k\in \mathds{T}}$, the operation $\mathcal{L}$ should be understood in the element-wise sense.

Note that, in a real experiment, the Allan variance of a clock reading deviation must be plotted for a set of measurement data with measurement noise, such as $y[k]$ in \eqref{eq:Csensm}.
In this paper, however, we analyze the Allan variance of the clock reading deviation without measurement noise, such as $h[k]$, for a theoretical evaluation of time scale generation, while a set of noisy measurement data is used in the time scale generation algorithm.

\subsubsection{Illustrative Problems}\label{sec:illnum}

Consider an ensemble of 10 independent atomic clocks.
The standard deviations of white frequency noise and random walk frequency noise are listed in Table~\ref{table:noisepara}, which are determined based on our hardware experiments.
The standard deviations of measurement noise are listed in Table~\ref{table:noiseparam}, where the phase differences are measured according to \eqref{eq:mesV}.

\begin{figure}[t]
\centering
\includegraphics[width = .99\linewidth]{ 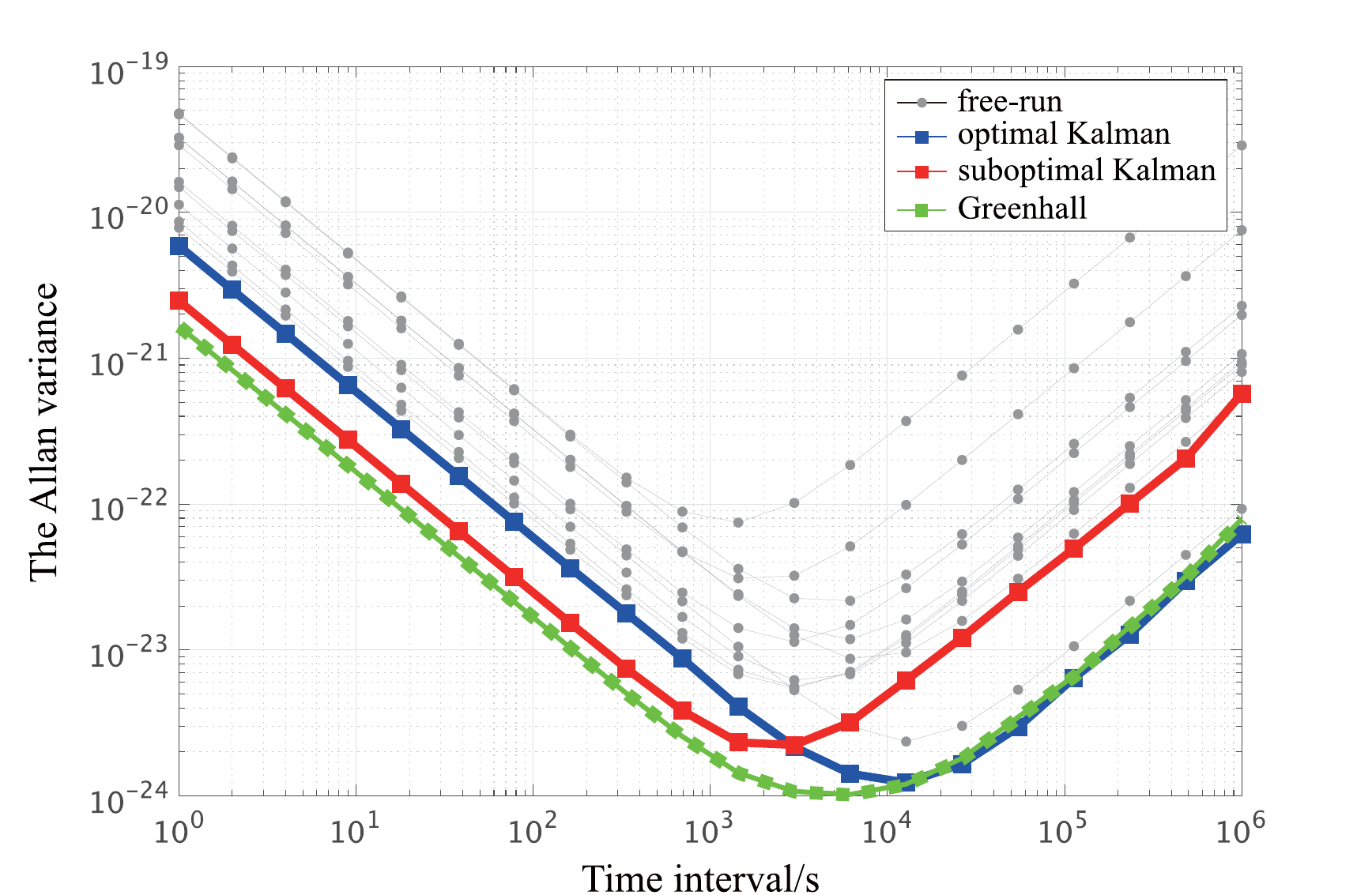}
\medskip
\caption{The Allan variances.
The grey lines are the analytical Allan variance of the free-running clocks.
The blue line is the statistical Allan variance of the reference time scale generated by the optimal Kalman filtering.
The red line is the statistical Allan variance of the reference time scale generated by the suboptimal Kalman filtering.
}
\label{fig:Kalallan}
\end{figure}

The analytical Allan variance in \eqref{eq:anAllan} is plotted by the grey lines for each clock in Fig.~\ref{fig:Kalallan}.
To this ensemble clock, we run the Kalman filtering algorithm in \eqref{eq:kalfil} to generate a reference time scale.
Specifically, the Kalman filtering is implemented for the stochastic time evolution of the ensemble clock with the time length $T$ of $10^{7}$~[s] and the sampling interval $\tau$ of 1~[s].
Then, the statistical Allan variance of the generated reference time scale $\mathcal{L}\{\varepsilon[k]\}_{k\in \mathds{T}}$ with
\[
\varepsilon[k] := (C \otimes \tfrac{1}{N} \mathds{1}_{N}^{\sf T}) e[k],
\]
where $e[k]$ is the state estimation error by the Kalman filtering, is plotted by the blue line in Fig.~\ref{fig:Kalallan}.
We can see that the statistical Allan variance of the reference time scale is better than the analytical Allan variance of each clock in both the short and long term.

As an illustration, using the same time sequence data, we show the result when the value of the noise covariance used in the Kalman filtering algorithm is artificially varied.
In particular, in the recursion of \eqref{eq:kalfil}, we use
\[
\bm{Q} = \mat{
(\tau \overline{\sigma}_1^2 + \frac{\tau^3}{3} \overline{\sigma}_2^2 ) I_{10}  & \frac{\tau^2}{2}\overline{\sigma}_2^2 I_{10} \\
\frac{\tau^2}{2} \overline{\sigma}_2^2 I_{10} & \tau \overline{\sigma}_2^2 I_{10}
}
\]
where $\overline{\sigma}_i^2$ denotes the average of noise variances
\[
\overline{\sigma}_i^2 := \frac{1}{10} \sum_{j=1}^{10}\sigma_{ij}^2 .
\]
Obviously, this setting is not optimal in terms of the state estimation via the Kalman filtering due to the modeling error of the noise variances.
The result is shown by the red line in Fig.~\ref{fig:Kalallan}.
Notably, the result for this suboptimal case is better than that for the optimal case in terms of short-term stability.

\begin{figure}[t]
\centering
\includegraphics[width = .99\linewidth]{ 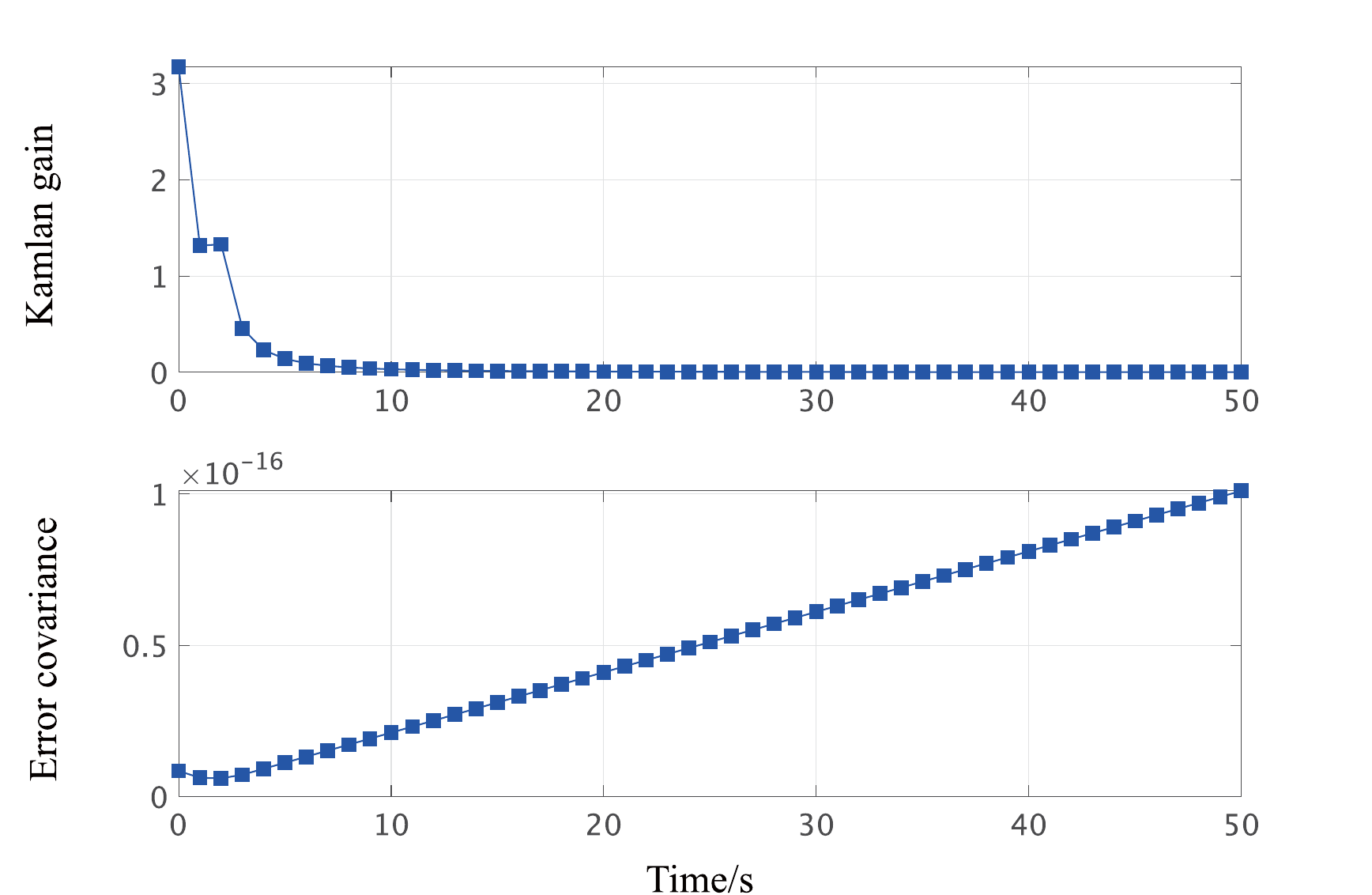}
\medskip
\caption{Increments of the Kalman gain and error covariance.
The upper subfigure shows the Frobenius norm of the increment of the Kalman gain.
The lower subfigure shows that of the error covariance.
}
\label{fig:KalPH}
\end{figure}

As shown above, the optimal Kalman filtering is generally not optimal for time scale generation in terms of the Allan variance.
In addition, this Kalman filtering also has the problem that the error covariance for calculating the Kalman gain diverges over time.
This is demonstrated in Fig.~\ref{fig:KalPH}, where the Frobenius norms \cite{bernstein2009matrix} of the increments, i.e.,
$\|\triangle_1 H[k]\|_{\rm F}$ and $\|\triangle_1 P^-[k]\|_{\rm F}$, are plotted.
Note that even though the Kalman gain is calculated by the product of the error covariance and a constant matrix, it converges to a constant matrix over time.
This implies that the Kalman filtering algorithm for ensemble clocks contains ``indeterminate'' operations, such that a diverging value is annihilated by a matrix multiplication.

For reference, we also show the results of the Kalman filtering modified with the KPW algorithm \cite{greenhall2006kalman} as the green line in Fig.~\ref{fig:Kalallan}. 
While the long-term stability is comparable to the optimal Kalman filtering in blue, the modified Kalman filtering demonstrates greater short-term stability.
Note that these results are all supposed to generate time scales using digital clock reading adjustments rather than physical oscillator frequency regulations.

From these observations, we ask the following questions.
\begin{itemize}
\item[{\bf Q1}]  Is it possible to find a Kalman filtering algorithm in a form that does not contain indeterminate operations? 
\item[{\bf Q2}]  Is it possible to find a control algorithm that synchronizes all the clocks to a user-specified time scale through physical oscillator frequency regulation rather than digital clock reading adjustment?
\item[{\bf Q3}]  Is it possible to mathematically prove why the optimal Kalman filtering algorithm does not produce the optimal time scale in terms of the Allan variance?
\item[{\bf Q4}]  Is it possible to generate an optimal time scale based on the Kalman filtering that achieves both short-term and long-term frequency stability?
\end{itemize}

In this paper, we give clear solutions to all these questions based on systems and control theory.
In particular, we develop a general framework, called explicit ensemble mean (EEM) synchronization, that unifies time scale generation, clock synchronization, and oscillator frequency regulation.

The following is a roadmap for asking the questions.
First, Section~3 analyzes the Kalman filtering algorithm for ensemble clocks based on the notion of the observable canonical decomposition \cite{antoulas2005approximation}.
In particular, 
\begin{itemize}
\item Section~3.1 derives all observable canonical decompositions of the ensemble clock model, separating the dynamics into observable and unobservable components that represent synchronization error and synchronization destination, respectively, 
\item Section~3.2 reformulates the standard Kalman filtering algorithm in the decomposed basis, and derives an equivalent determinate form that removes the indeterminate operations, and
\item Section~3.3 analyzes the stationary version of the determinate Kalman filtering, and derives an explicit expression for the stationary Kalman gains. 
\end{itemize}
Then, Section~4 develops the proposed EEM synchronization framework.
In particular,
\begin{itemize}
\item Section~4.1 introduces the EEM synchronization framework by selecting a basis in which the unobservable state explicitly represents the ensemble mean with a user-specified weight, and derives a clock synchronization algorithm associated with the ensemble mean weight,
\item Section 4.2 evaluates the synchronization destination in terms of the Allan variance, and derives the optimal ensemble mean weights that minimize the Allan variance in the short and long terms, and
\item Section 4.3 shows that the Kalman filtering for time scale generation can be understood as a special
case of the EEM synchronization with the best long-term stability, and develops a control strategy that balances short-term and long-term stability.
\end{itemize}


\section{Analysis of Kalman Filtering Algorithm}

\subsection{All Observable Canonical Decompositions}

We apply observable canonical decomposition \cite{antoulas2005approximation} to the ensemble clock model in \eqref{eq:Csensm} for the analysis of the Kalman filtering algorithm in \eqref{eq:kalfil}.
For this purpose, we consider a basis expansion based on the unobservable subspace as
\begin{equation}\label{eq:obscan}
x[k] = \bm{W} \bigl( (I_2 \otimes V)\bm{W} \bigr)^{-1} \xi_o[k]
+ (I_2 \otimes \mathds{1}_N)\xi_{\bar{o}}[k]
\end{equation}
where $\xi_o[k]$ is the observable state, $\xi_{\bar{o}}[k]$ is the unobservable state, and  $\bm{W}\in \mathbb{R}^{2N \times 2(N-1)}$ is a column full rank matrix that represents a parameter of the decomposition such that
\begin{equation}\label{eq:allR2N}
\sfim \bm{W} + \sfim (I_2 \otimes \mathds{1}_N) = \mathbb{R}^{2N}.
\end{equation}
Owing to \eqref{eq:allR2N}, for the new state vector
\[
\xi[k]:= \mat{
\xi_o[k] \\
\xi_{\bar{o}}[k]
},
\]
we have the one-to-one relationship as
\[
\xi [k]
=
\bm{T} x[k] 
\quad
\iff
\quad
x[k]  = \bm{T}^{-1} \xi [k]
\]
with the transformation matrices
\begin{equation}\label{eq:bmTc}
\spliteq{
\bm{T}^{-1} &= \mat{
\bm{W} \bigl( (I_2 \otimes V)\bm{W} \bigr)^{-1} &
I_2 \otimes \mathds{1}_N
}, \\
\bm{T}&=
\mat{
I_2 \otimes V \\
\bigl( \overline{\bm{W}} (I_2 \otimes \mathds{1}_N) \bigr)^{-1} \overline{\bm{W}}
}
}
\end{equation}
where $\overline{\bm{W}} \in \mathbb{R}^{2\times 2N}$ is a row full rank matrix such that
\begin{equation}\label{eq:WbWcon}
\sfker \overline{\bm{W}} = \sfim \bm{W}.
\end{equation}
This relationship implies that choosing the column space of $\bm{W}$ is equivalent to choosing the row space of $\overline{\bm{W}}$.
Note that $\xi_{\bar{o}}[k]$ in \eqref{eq:obscan} represents the unobservable state of the ensemble clock model because the unobservable subspace is given as in \eqref{eq:unobsss}.
On the other hand, $\xi[k]$ represents the observable state whose basis is parameterized by $\bm{W}$.

Applying this basis expansion, or variable change, we have the new state-space realization as
\begin{equation}\label{eq:newss}
\simode{
\xi [k+1]
&= 
\bm{T}\bm{A}\bm{T}^{-1}
\xi [k]  +
\bm{T}\bm{B}
u[k]
+
\bm{T}
v[k] \\
y[k] &= \bm{C}\bm{T}^{-1} \xi [k]
+w[k].
}
\end{equation}
Note that the new system matrices have the structure of
\[
\spliteq{
\tilde{\bm{A}} &:=\bm{T}\bm{A}\bm{T}^{-1}=
\mat{
\bm{A}_o & 0 \\
\overline{\bm{U}} \bm{A} \bm{U} & A
},
\\
\tilde{\bm{B}} &:=\bm{T}\bm{B}=
\mat{
\bm{B}_o V \\
\overline{\bm{U}}
\bm{B}
},  
\\
\tilde{\bm{C}} &:= \bm{C}\bm{T}^{-1}=
\mat{
\bm{C}_o & 0
}
}
\]
where the generalized inverses are represented as
\[
\bm{U}:=
\bm{W} \bigl( (I_2 \otimes V)\bm{W} \bigr)^{-1}
,\quad
\overline{\bm{U}}:=
\bigl( \overline{\bm{W}} (I_2 \otimes \mathds{1}_N) \bigr)^{-1} \overline{\bm{W}},
\]
and the system matrices of the observable subsystem as
\[
\bm{A}_o := A \otimes I_{N-1},\quad
\bm{B}_o := B \otimes I_{N-1},\quad
\bm{C}_o := C \otimes I_{N-1}.
\]
It is clear from the block triangular structure that the unobservable state $\xi_{\bar{o}}[k]$ has no effect on the measurement output $y[k]$.
This is a complete parameterization of the observable canonical decomposition of the ensemble clock model.

\subsection{Determinate Kalman Filtering Algorithm}

Consider the Kalman filtering algorithm for the new state-space realization in \eqref{eq:newss} as
\begin{equation}\label{eq:kalfil2}
\simode{
\hat{\xi }^-[k] &= \tilde{\bm{A}} \hat{\xi }[k-1] + \tilde{\bm{B}} u[k-1] \\
\tilde{\bm{P}}^- [k] &= \tilde{\bm{A}} \tilde{P}[k-1] \tilde{\bm{A}}^{\sf T} + \tilde{\bm{Q}}\\
\tilde{\bm{H}}[k] &= \tilde{\bm{P}}^-[k] \tilde{\bm{C}}^{\sf T} \bigl( \tilde{\bm{C}} \tilde{\bm{P}}^-[k] \tilde{\bm{C}}^{\sf T} + \bm{R} \bigr)^{-1} \\
\tilde{\bm{P}}[k] &= \bigl( I_{2N} - \tilde{\bm{H}}[k] \tilde{\bm{C}} \bigr) \tilde{\bm{P}}^-[k] \\
\hat{\xi }[k] &= \hat{\xi }^-[k] + \tilde{\bm{H}}[k] \bigl(y[k] - \tilde{\bm{C}} \hat{\xi }^-[k] \bigr) 
}
\end{equation}
where the symbols with the tilde are new variables.
In fact, this Kalman filtering algorithm is equivalent to the original one in \eqref{eq:kalfil} because they have the one-to-one relationship as
\[
\spliteq{
\hat{\xi }^-[k] &= \bm{T} \hat{x}^-[k]
,\quad
\hat{\xi }[k] = \bm{T} \hat{x}[k]
,\quad
\tilde{\bm{H}}[k] = \bm{T} \bm{H}[k] \\
\tilde{\bm{P}}^-[k] &= \bm{T} \bm{P}^-[k] \bm{T}^{\sf T}
,\quad
\tilde{\bm{P}}[k] = \bm{T} \bm{P}[k] \bm{T}^{\sf T}
,\quad
\tilde{\bm{Q}} = \bm{T} \bm{Q} \bm{T}^{\sf T} .
}
\]
For the purposes of the following discussion, we denote the estimation error covariances by
\[
\tilde{\bm{P}}^-[k] = 
\mat{
\bm{P}^-_{oo}[k] & * \\
P^-_{\bar{o}o}[k] & P^-_{\bar{o}\bar{o}}[k]
}
,\quad
\tilde{\bm{P}}[k] = 
\mat{
\bm{P}_{oo}[k] & * \\
P_{\bar{o}o}[k] & P_{\bar{o}\bar{o}}[k]
},
\]
where the symmetric parts are omitted.
Note that 
$\bm{P}_{oo}[k]$ represents the covariance of the observable state estimation error, 
$P_{\bar{o}\bar{o}}[k]$ represents that of the unobservable state estimation error, and
$P_{\bar{o}o}[k]$ represents that of the unobservable and observable state estimation errors.
Then, we see that
\[
\tilde{\bm{P}}^- [k]\tilde{\bm{C}}^{\sf T} = \tilde{\bm{A}} \tilde{\bm{P}}[k-1] \tilde{\bm{A}}^{\sf T}\tilde{\bm{C}}^{\sf T}
 + \tilde{\bm{Q}}\tilde{\bm{C}}^{\sf T}\\
\]
can be represented as
\[
\mat{
\bm{P}^-_{oo}[k]  \\
P^-_{\bar{o}o}[k] 
}
\bm{C}_o^{\sf T}
 =
\left(
\tilde{\bm{A}}
\mat{
\bm{P}_{oo}[k-1]  \\
P_{\bar{o}o}[k-1] 
}
\bm{A}_o^{\sf T}
+
\mat{
\bm{Q}_o \\
Q_{\bar{o}} 
}
\right)
\bm{C}_o^{\sf T},
\]
where the process noise covariances are defined as
\[
\bm{Q}_o := (I_2 \otimes V) \bm{Q} (I_2 \otimes V)^{\sf T}
,\quad
Q_{\bar{o}} := \overline{\bm{U}} \bm{Q} (I_2 \otimes V)^{\sf T}.
\]
This means that the update equation of $\tilde{\bm{P}}^-[k]$ is independent of $P^-_{\bar{o}\bar{o}}[k]$ and $P_{\bar{o}\bar{o}}[k]$, and it is given as 
\[
\mat{\bm{P}^-_{oo} [k]\\ P^-_{\bar{o}o} [k]} = 
\mat{
\bm{A}_o & 0 \\
\overline{\bm{U}} \bm{A} \bm{U} & A
}
\mat{
\bm{P}_{oo} [k\!-\!1] \\
P_{\bar{o}o} [k\!-\!1]
}
\bm{A}_o^{\sf T}
+
\mat{
\bm{Q}_o \\
Q_{\bar{o}} 
}.
\]
Similarly, the Kalman gain
\[
\tilde{\bm{H}}[k] = \mat{\bm{H}_o[k] \\ H_{\bar{o}} [k] }
\]
is also independent of $P^-_{\bar{o}\bar{o}}[k]$ and $P_{\bar{o}\bar{o}}[k]$ because
\[
\mat{\bm{H}_o[k] \\ H_{\bar{o}} [k] } = 
\mat{
\bm{P}^-_{oo}[k]  \\ P^-_{\bar{o}o}[k] 
}
\bm{C}_o^{\sf T}
\bigl( \bm{C}_o \bm{P}^-_{oo}[k] \bm{C}_o^{\sf T} + \bm{R} \bigr)^{-1}.
\]
Finally, considering the symmetry of the error covariances, we have the update equations of $\bm{P}_{oo} [k]$ and $P_{\bar{o}o} [k]$ as
\[
\mat{\bm{P}_{oo} [k]\\ P_{\bar{o}o} [k]} = 
\mat{ \bigl( I_{2(N-1)} - \bm{H}_o[k] \bm{C}_o  \bigr) \bm{P}^-_{oo}[k] \\
P^-_{\bar{o}o}[k] \bigl( I_{2(N-1)} - \bm{C}_o^{\sf T} \bm{H}_o^{\sf T}[k] \bigr) 
}.
\]
Note that $P_{\bar{o}\bar{o}}$, which is the cause of the indeterminate operations as it diverges over time in the standard Kalman filtering algorithm, is removed from the update equations.
In summary, the following Kalman filtering algorithm without indeterminate operations is obtained.

\begin{lemma}\label{lem:decKal}
The Kalman filtering algorithm in \eqref{eq:kalfil} and its determinate version given as
\begin{equation}\label{eq:decKalman}
\simode{
\mat{
\hat{\xi}^-_o[k] \\
\hat{\xi}_{\bar{o}}^-[k]
}
&= 
\mat{
\bm{A}_o & 0 \\
\overline{\bm{U}} \bm{A} \bm{U} & A
}
\mat{
\hat{\xi}_o[k\!-\!1] \\
\hat{\xi}_{\bar{o}}[k\!-\!1]
}
\!+\!
\mat{
\bm{B}_o V \\
\overline{\bm{U}}
\bm{B}
}
\! u[k\!-\!1]   
\\
\mat{\bm{P}^-_{oo} [k]\\ P^-_{\bar{o}o} [k]} &= 
\mat{
\bm{A}_o & 0 \\
\overline{\bm{U}} \bm{A} \bm{U} & A
}
\mat{
\bm{P}_{oo} [k\!-\!1] \\
P_{\bar{o}o} [k\!-\!1]
}
\bm{A}_o^{\sf T}
+
\mat{
\bm{Q}_o \\
Q_{\bar{o}} 
}
\\
\mat{\bm{H}_o[k] \\ H_{\bar{o}} [k] } &= 
\mat{
\bm{P}^-_{oo}[k]  \\ P^-_{\bar{o}o}[k] 
}
\bm{C}_o^{\sf T}
\bigl( \bm{C}_o \bm{P}^-_{oo}[k] \bm{C}_o^{\sf T} + \bm{R} \bigr)^{-1}
\\
\mat{\bm{P}_{oo} [k]\\ P_{\bar{o}o} [k]} &= 
\mat{ \bigl( I_{2(N-1)} - \bm{H}_o[k] \bm{C}_o  \bigr) \bm{P}^-_{oo}[k] \\
P^-_{\bar{o}o}[k] \bigl( I_{2(N-1)} - \bm{C}_o^{\sf T} \bm{H}_o^{\sf T}[k] \bigr) 
}
 \\
 \mat{
 \hat{\xi}_o[k] \\ \hat{\xi}_{\bar{o}}[k]
 }
 &= \mat{
\hat{\xi}_o^-[k] \\
\hat{\xi}_{\bar{o}}^-[k]
} + \mat{\bm{H}_o[k] \\ H_{\bar{o}} [k] } 
\bigl(y[k] - \bm{C}_o \hat{\xi}_o^-[k] \bigr) 
}
\end{equation}
are equivalent such that
\[
\hat{x}[k] =
\bm{U} \hat{\xi}_o[k]
+
\bigl( I_2 \otimes \mathds{1}_N \bigr) \hat{\xi}_{\bar{o}}[k]
,\quad
k\in \mathbb{Z}
\]
for any basis parameter $\bm{W}$, or equivalently $\overline{\bm{W}}$.
\end{lemma}


Lemma~\ref{lem:decKal} shows that indeterminate operations involved in the standard Kalman filtering algorithm can be removed in an equivalent way.
In fact, the stationary values of the error covariances for the determinate equivalent in \eqref{eq:decKalman} can be found, while the error covariances for the original algorithm diverge over time.

It should be emphasized that, even if either $\overline{\bm{U}} \bm{A} \bm{U}$ or $Q_{\bar{o}}$  is zero, the estimation of the unobservable state $\xi_{\bar{o}}[k]$ is necessary for minimizing the state estimation error.
In general, the estimate $\hat{\xi}_{\bar{o}}[k]$ has a non-zero value because the Kalman gain $H_{\bar{o}}[k]$ is not zero in the unobservable part.
The estimation of the the unobservable state will play an important role in generating an optimal time scale.


\subsection{Determinate Stationary Kalman Filtering Algorithm}

Unlike the original algorithm, the determinate Kalman filtering algorithm converges to a stationary algorithm over time.
In time scale generation, the properties of such a stationary algorithm are inherently important because clock state estimation continues over a long period of time.
To illustrate, we denote the stationary error covariances by
\[
\bm{P}_{oo}^{\star} :=\lim_{k\rightarrow \infty} \bm{P}_{oo}^- [k]
,\quad
P_{\bar{o}o}^{\star} :=\lim_{k\rightarrow \infty} P_{\bar{o}o}^- [k],
\]
and the stationary Kalman gains by
\[
\bm{H}_{o}^{\star} :=\lim_{k\rightarrow \infty} H_{o} [k]
,\quad
H_{\bar{o}}^{\star} :=\lim_{k\rightarrow \infty} H_{\bar{o}} [k].
\]
Then, the stationary algorithm is found as follows.

\begin{theorem}\label{thm:decKalst}
For the determinate Kalman filtering algorithm in \eqref{eq:decKalman}, the stationary error covariances are given as the unique solutions to the algebraic equations
\begin{equation}\label{eq:Pricc}
\spliteq{
\bm{P}_{oo}^{\star} &= \bm{Q}_o + 
\bm{A}_o \bm{P}_{oo}^{\star} 
\bm{S}(\bm{P}^{\star}_{oo})
\bm{A}_o^{\sf T},
\\
P_{\bar{o}o}^{\star} &= Q_{\bar o} +
\bigl( A P_{\bar{o}o}^{\star} + \overline{\bm{U}} \bm{A} \bm{U} \bm{P}^{\star}_{oo} \bigr)
\bm{S}(\bm{P}^{\star}_{oo}) \bm{A}_o^{\sf T} 
}
\end{equation}
where $\bm{S} (\bm{P}^{\star}_{oo})$ is defined as
\[
\bm{S} (\bm{P}^{\star}_{oo}):=
I_{2(N-1)} - \bm{C}_o^{\sf T} ( \bm{C}_o \bm{P}^{\star}_{oo} \bm{C}_o^{\sf T} + \bm{R} )^{-1} 
\bm{C}_o \bm{P}^{\star}_{oo}.
\]
Furthermore, the stationary algorithm is given as
\begin{equation}\label{eq:decKFst}
\simode{
\mat{
\hat{\xi}^-_o[k] \\
\hat{\xi}_{\bar{o}}^-[k]
}
&= 
\mat{
\bm{A}_o & 0 \\
\overline{\bm{U}} \bm{A} \bm{U} & A
}
\mat{
\hat{\xi}_o[k\!-\!1] \\
\hat{\xi}_{\bar{o}}[k\!-\!1]
}
\!+\!
\mat{
\bm{B}_o V \\
\overline{\bm{U}}
\bm{B}
}
\! u[k\!-\!1]   
\\
\mat{
 \hat{\xi}_o[k] \\ \hat{\xi}_{\bar{o}}[k]
 }
 &= \mat{
\hat{\xi}_o^-[k] \\
\hat{\xi}_{\bar{o}}^-[k]
} + \mat{\bm{H}_o^{\star} \\ H_{\bar{o}}^{\star} } 
\bigl(y[k] - \bm{C}_o \hat{\xi}_o^-[k] \bigr) 
}
\end{equation}
where the stationary Kalman gains are given as
\begin{equation}\label{eq:stkalgain}
\spliteq{
\bm{H}_{o}^{\star} & = \bm{P}_{oo}^{\star} \bm{C}_o^{\sf T}
\bigl( \bm{C}_o \bm{P}_{oo}^{\star} \bm{C}_o^{\sf T} + \bm{R} \bigr)^{-1}, \\
H_{\bar{o}}^{\star} & = P_{\bar{o}o}^{\star} \bm{C}_o^{\sf T}
\bigl( \bm{C}_o \bm{P}_{oo}^{\star} \bm{C}_o^{\sf T} + \bm{R} \bigr)^{-1}.
}
\end{equation}
\end{theorem}

\begin{IEEEproof}
By considering the recursive update equations in \eqref{eq:decKalman} under the conditions that 
\[
\bm{P}^-_{oo}[k] = \bm{P}^-_{oo}[k+1] = \bm{P}^{\star}_{oo}
,\quad
P^-_{\bar{o}o}[k] = P^-_{\bar{o}o}[k+1] = P^{\star}_{\bar{o}o},
\]
we have the algebraic equations for $\bm{P}^{\star}_{oo}$ and $P^{\star}_{\bar{o}o}$ as in \eqref{eq:Pricc}.
The representation of the stationary Kalman gains in \eqref{eq:stkalgain} are obtained from their update equations.


Next, we prove that $\bm{P}_{oo}^{\star} $ and $P_{\bar{o}o}^{\star} $ are uniquely determined.
The uniqueness of $\bm{P}_{oo}^{\star} $ is proven because the Kalman filtering algorithm for the observable part coincides with the standard algorithm.
On the other hand, to prove the uniqueness of $P_{\bar{o}o}^{\star} $, we use the fact that
\[
\bigl(
I_{2(N-1)} - \bm{H}_o^{\star} \bm{C}_o 
\bigr)^{\sf T}= 
\bigl(
I_{2(N-1)} + \bm{C}_o^{\sf T} R^{-1} \bm{C}_o \bm{P}^{\star}_{oo}
\bigr)^{-1}.
\]
This is proven based on the fact that
\[
\bm{S}(\bm{P}^{\star}_{oo})=
\bigl(
I_{2(N-1)} + \bm{C}_o^{\sf T} R^{-1} \bm{C}_o \bm{P}^{\star}_{oo}
\bigr)^{-1},
\]
which is known as the matrix inversion lemma.
Then, the equation of $P_{\bar{o}o}^{\star} $ in \eqref{eq:Pricc} is written as
\[
\spliteq{
P_{\bar{o}o}^{\star} &= 
A P_{\bar{o}o}^{\star} 
\bigl(
I_{2(N-1)} - \bm{H}_o^{\star} \bm{C}_o 
\bigr)^{\sf T}
\bm{A}_o^{\sf T} \\
& + 
\underbrace{
	Q_{\bar o} + \overline{\bm{U}} \bm{A} \bm{U} \bm{P}^{\star}_{oo}
	\bigl(
	I_{2(N-1)} - \bm{H}_o^{\star} \bm{C}_o 
	\bigr)^{\sf T}
	 \bm{A}_o^{\sf T}
	 }_{X}.
}
\]
Using the vectorization of matrices defined as
\[
\sfvec \left(\mat{x_1 & \cdots & x_m} \right) = 
\mat{
x_1^{\sf T} & \cdots & x_m^{\sf T}
}^{\sf T},
\]
where $x_i$ is a vector, we have the linear equation
\[
\sfvec (P_{\bar{o}o}^{\star})
\!=\!
\bigl\{
	\bm{A}_o
	\bigl(
	I_{2(N-1)} \!-\! \bm{H}_o^{\star} \bm{C}_o 
	\bigr)
\otimes A
\bigr\}
\sfvec (P_{\bar{o}o}^{\star}) \!+\! \sfvec (X),
\]
which means that $P_{\bar{o}o}^{\star}$ is uniquely determined if 
\[
Y := I_{4(N-1)} - \bigl\{
	\bm{A}_o
	\bigl(
	I_{2(N-1)} \!-\! \bm{H}_o^{\star} \bm{C}_o 
	\bigr)
\otimes A
\bigr\}
\]
is nonsingular.
Because $\bm{H}_o^{\star}$ is the stationary Kalman gain for the observable part, we see that
\[
\bm{\Lambda} \left(
\bm{A}_o
\bigl(
I_{2(N-1)} \!-\! \bm{H}_o^{\star} \bm{C}_o 
\bigr)
\right) \subseteq \mathds{D}.
\]
Note that the eigenvalues of $A$ are one.
Thus, we have
\[
\bm{\Lambda} \left(
\bm{A}_o
\bigl(
I_{2(N-1)} \!-\! \bm{H}_o^{\star} \bm{C}_o 
\bigr) \otimes A
\right) \subseteq \mathds{D}.
\]
This proves that $Y$ is nonsingular.
\end{IEEEproof}
\medskip

The stationary Kalman filtering algorithm in Theorem~\ref{thm:decKalst} does not contain indeterminate operations, nor does it require recursive computation of error covariances.
This is a solution to {\bf Q1}.
We remark that there is no need to specify the initial error covariances, which also avoids unexpected behavior due to inappropriate initial values in practice.
Furthermore, $\bm{H}_o^{\star}$ can be computed as the standard Kalman gain for the observable state dynamics, while $H_{\bar{o}}^{\star}$ can easily be computed from $\bm{H}_o^{\star}$ without solving a linear matrix equation.
This will be shown later.

The following mathematical analysis is performed on the stationary version in \eqref{eq:decKFst}. 
For the actual implementation of clock state estimation, however, either the stationary or non-stationary version may be used, depending on the situation. 
For example, the non-stationary version in \eqref{eq:decKalman} may be more robust against non-stationary clock noise behavior.


\section{Explicit Ensemble Mean Synchronization}

\subsection{General Framework}

\subsubsection{Specific Choice of Basis Parameter }

In the rest of this paper, we consider the stationary Kalman filtering algorithm in \eqref{eq:decKFst} for the mathematical analysis of time scale generation.
The main idea is to choose the basis parameter as
\begin{equation}\label{eq:bmWbkr}
\overline{\bm{W}} = I_2 \otimes q^{\sf T}
\end{equation}
using the parameter vector $q\in \mathbb{R}^{N}$ to represent a user-specified ensemble mean weight of the clocks.
Without loss of generality, we can assume that
\begin{equation}\label{eq:normqkf}
q^{\sf T} \mathds{1}_N =1.
\end{equation}
This is equivalent to selecting
\[
\bm{W} = I_2 \otimes V^+
\]
where $V^+\in \mathbb{R}^{N\times (N-1)}$ is the generalized inverse of $V$ associated with $q$.
In particular, using a column full rank matrix $W \in \mathbb{R}^{N \times (N-1)}$ such that
\begin{equation}\label{eq:qWcon}
q^{\sf T} W =0,
\end{equation}
we can represent the generalized inverse as
\begin{equation}\label{eq:geninv}
V^+ := W (VW)^{-1} .
\end{equation}
We can easily verify that
\begin{equation}\label{eq:kfginv}
V V^+ = I_{N-1}
,\quad
q^{\sf T} V^+=0.
\end{equation}
Note that $W$ as well as $V^+$ contain information equivalent to $q$ because
\[
\sfim W = \sfim V^+ = \sfker q^{\sf T}.
\]
Furthermore, we see that
\[
\bm{W} = \bm{U}
,\quad
\overline{\bm{W}} = \overline{\bm{U}}.
\]
In this parameter choice, the basis expansion is found as
\begin{equation}\label{eq:kfstex}
x[k] = (I_2 \otimes V^+) \xi_o[k]
+ (I_2 \otimes \mathds{1}_N)\xi_{\bar{o}}[k].
\end{equation}
Therefore, we can see from \eqref{eq:normqkf} and \eqref{eq:kfginv} that the unobservable state given as
\[
\xi_{\bar{o}}[k]
= (I_2 \otimes q^{\sf T}) x[k]
\]
represents the weighted average or explicit ensemble mean of  clocks.
On the other hand, the observable state 
\[
\xi_o[k]
= (I_2 \otimes V) x[k]
\]
represents the relative clock state or synchronization error with respect to the ensemble mean.
In this basis parameter choice, we can verify that the observable state dynamics
\begin{subequations}\label{eq:qdecsys}
\begin{equation}\label{eq:obssubs}
\simode{
\xi_o[k +1] &= \bm{A}_o \xi_o[k] + (I_2 \otimes V)v[k]
+ \bm{B}_o \omega_o[k] \\
y[k] &= \bm{C}_o \xi_o[k] + w[k]
}
\end{equation}
does not affect the unobservable state dynamics 
\begin{equation}\label{eq:unobssubs}
\xi_{\bar{o}}[k+1]  = A \xi_{\bar{o}}[k] + (I_2 \otimes q^{\sf T}) v[k] 
+ B \omega_{\bar{o}}[k]  
\end{equation}
where the control input is expanded as
\begin{equation}\label{eq:uexpand}
u[k] = V^{+} \omega_o[k] 
+ \mathds{1}_N \omega_{\bar{o}}[k].
\end{equation}
\end{subequations}
Note that $\omega_o[k]$ is the control input to achieve clock synchronization, while $\omega_{\bar{o}}[k]$ is the control input to regulate synchronization destination.
In the following, the former is called the synchronization control input, and the latter is called the collective control input.

The generalized inverse $V^+$ in \eqref{eq:geninv} is not dependent on the specific choice of $W$, but is dependent on $q$ and $V$.
Note that it may not be possible to find a simple expression for $V^+$ given a general choice of $V$.
As a special case, for the choice of $V$ in \eqref{eq:mesV},  we can find the analytical expression
\[
V^+ = \mat{I_{N-1}\\0} - \mathds{1}_{N} \mathds{1}_{N-1}^{\sf T} \sfdiag(q_i)_{i\in \mathds{N}^-},
\]
which satisfies \eqref{eq:kfginv}.
This means that the state of each clock is expanded as
\[
x_i[k] = \xi_{oi} [k] - \sum_{j\in \mathds{N}^-} q_j \xi_{oj} [k]
+ \xi_{\bar{o}}[k]
\]
for $i \in \mathds{N}^-$  and
\[
x_N[k] = - \sum_{j\in \mathds{N}^-} q_j \xi_{oj} [k]
+ \xi_{\bar{o}}[k],
\]
where $\xi_{oi} [k]$ is the $i$th element of the observable state $\xi_{o} [k]$, which is equal to the relative clock state $\Delta x_i^N[k]$ in \eqref{eq:relst}.
The unobservable state is the weighted average of all clocks as
\[
\xi_{\bar{o}}[k] = \sum_{j\in \mathds{N}} q_j x_j[k].
\]
In a similar way, the control input to each clock can be expressed as
\[
u_i[k] = \omega_{oi} [k] - \sum_{j\in \mathds{N}^-} q_j \omega_{oj} [k]
+ \omega_{\bar{o}}[k]
\]
for $i \in \mathds{N}^-$  and
\[
u_N[k] = - \sum_{j\in \mathds{N}^-} q_j \omega_{oj} [k]
+ \omega_{\bar{o}}[k],
\]
where $\omega_{oi} [k]$ is the $i$th element of the synchronization control input $\omega_{o} [k]$.

The main advantage of the basis expansion in \eqref{eq:kfstex} is that the control effort to keep the magnitude of the observable state $\xi_o[k]$ small implies synchronizing the clock states with their ensemble mean, or equivalently the unobservable state $\xi_{\bar{o}}[k]$.
This implies that ``explicit" ensemble mean synchronization can be achieved with a user-specified weight $q$ for the clocks.
Because the observable state dynamics in \eqref{eq:obssubs} is both observable and controllable, it is easy to find a control algorithm that constructs a synchronization control input $\omega_o[k]$ to stabilize the observable state dynamics.
This will be explained in the next subsection.

The column space of $I_2 \otimes V^+$, which is dependent on $q$, is used for the basis of the observable state $\xi_o[k]$ in \eqref{eq:kfstex}.
Note that the observable state dynamics in \eqref{eq:obssubs} does not depend on the choice of $q$.
Conversely, the basis of the unobservable state $\xi_{\bar{o}}[k]$, which is spanned by the column space of $I_2 \otimes \mathds{1}_N$, does not depend on $q$, but the unobservable state dynamics in \eqref{eq:unobssubs} does depend on $q$.
From this observation, we can see that selecting the basis of the observable state $\xi_o[k]$ as the column space of $I_2 \otimes V^+$ is more important than selecting the component $\xi_o[k]$ as the relative clock state. 
Such a particular ``basis" selection of the observable state allows the unobservable state to represent the explicit ensemble mean.

\subsubsection{Explicit Ensemble Mean Synchronization Algorithm}

In the following, we analyze the stochastic characteristics of the EEM synchronization algorithm.
Because both process noise and measurement noise are assumed to be white and Gaussian, and the system of difference equations is linear, all time sequences of stochastic variables, which are weighted sums of Gaussian processes, are also Gaussian processes.

We examine a specific algorithm for achieving the EEM synchronization.
The basic strategy combines state estimation and state feedback control.
In particular, using the Kalman filtering algorithm, we apply 
\begin{subequations}\label{eq:Kalent}
\begin{equation}\label{eq:Kalobs}
\simode{
\hat{\xi}_o^-[k+1] &= \bm{A}_o \hat{\xi}_o^-[k] \! + \! \bm{B}_o \omega_o[k] 
\!+ \! \bm{A}_o\bm{H}_o^{\star} 
\bigl(y[k] \!-\! \bm{C}_o \hat{\xi}_o^-[k] \bigr)\\
\omega_o[k] &= - \bm{F}_o \hat{\xi}_o^-[k]
}
\end{equation}
for clock synchronization, and
\begin{equation}\label{eq:Kalunobs}
\simode{
\hat{\xi}_{\bar{o}}^-[k+1] &= A \hat{\xi}_{\bar{o}}^-[k] + B \omega_{\bar{o}}[k] 
+ AH_{\bar{o}}^{\star}
\bigl(y[k] - \bm{C}_o \hat{\xi}^-_o[k] \bigr)\\
\omega_{\bar{o}}[k] &= - F_{\bar{o}}[k] \hat{\xi}_{\bar{o}}^-[k]
}
\end{equation}
\end{subequations}
for the regulation of synchronization destination.
The state estimation part is identical to the determinate Kalman filtering algorithm in \eqref{eq:decKFst} with the basis parameter in \eqref{eq:bmWbkr}.
The feedback gain $\bm{F}_o$ for the observable part is chosen such that
\begin{equation}\label{eq:obsFst}
\bm{\Lambda} (\bm{A}_o - \bm{B}_o \bm{F}_o) \subseteq \mathds{D},
\end{equation}
which is equivalent to the condition that the observable state, or its estimate, is stabilized in the sense of
\begin{equation}\label{eq:obsFsteq}
\{\xi_o[k]\}_{k\in \mathbb{Z}} \in \mathcal{G}
\quad
\iff
\quad
\{\hat{\xi}^-_o[k]\}_{k\in \mathbb{Z}} \in \mathcal{G},
\end{equation}
the equivalence of which follows from
\[
\{ \xi_o[k] - \hat{\xi}^-_o[k]\}_{k\in \mathbb{Z}} \in \mathcal{G}
\]
by the Kalman filtering.
On the other hand, the feedback gain $F_{\bar{o}}[k]$ for the unobservable part will be specified later.

To analyze the synchronization destination, we consider the free-running ensemble mean dynamics
\begin{equation}\label{eq:syncdes}
\Pi(q):
\simode{
r[k+1] & = A r[k] + (I_2 \otimes q^{\sf T}) v[k] \\
z[k] &= C r[k].
}
\end{equation}
The synchronization error centered on this synchronization destination is defined as
\begin{equation}\label{eq:deltak}
\delta[k]:= x[k] - (I_2 \otimes \mathds{1}_N) r[k].
\end{equation}
Then, clock synchronization can be achieved as follows.

\begin{theorem}\label{thm:synccon}
For the ensemble clock model in \eqref{eq:Csensm}, consider the observable canonical decomposition in \eqref{eq:qdecsys}.
Suppose that the feedback gain $F_{\bar{o}}[k]$ is zero.
Then, the synchronization control algorithm in \eqref{eq:Kalent} achieves
\begin{equation}\label{eq:xsync}
\{\delta[k]\}_{k\in \mathbb{Z}} \in \mathcal{G}
\end{equation}
for the free-running ensemble mean dynamics $\Pi(q)$ in \eqref{eq:syncdes} 
if and only if the feedback gain $\bm{F}_o$ is given as in \eqref{eq:obsFst}.
\end{theorem}

\begin{IEEEproof}
Because $F_{\bar{o}}[k]$ is supposed to be zero, the control input $\omega_{\bar{o}}[k]$ is zero for the unobservable state dynamics in \eqref{eq:unobssubs}.
Therefore, the dynamics of $\Pi(q)$ in \eqref{eq:syncdes} is identical to that of \eqref{eq:unobssubs}.
From the relationship in \eqref{eq:kfstex}, we see that
\[
\delta[k] = (I_2 \otimes V^+)\xi_o[k].
\]
Therefore, \eqref{eq:obsFst}, which is equivalent to \eqref{eq:obsFsteq}, implies \eqref{eq:xsync}.
On the other hand, we have
\[
(I_2 \otimes V) \delta[k] = \xi_o[k].
\]
Therefore, \eqref{eq:xsync} implies \eqref{eq:obsFst}.
\end{IEEEproof}
\medskip

Theorem~\ref{thm:synccon} states that all clocks synchronize with the free-running ensemble mean dynamics $\Pi(q)$ in \eqref{eq:syncdes} if and only if the observable state dynamics is stabilized.
Note that a desirable feedback gain $\bm{F}_o$ such that \eqref{eq:obsFst} holds can be easily found by a standard technique, because the observable state dynamics in \eqref{eq:obssubs} is both controllable and observable; see \cite[Chapters~2 and 5]{fairman1998linear} for example. 
This is a solution to {\bf Q2}.

The observable state dynamics in \eqref{eq:obssubs} does not depend on the ensemble mean weight $q$.
This means that estimating the observable state $\xi_o[k]$ and constructing the synchronization control input $\omega_o[k]$ can be done independently of the choice of $q$.
In contrast, the control input $u[k]$ in \eqref{eq:uexpand} does depend on $q$ because the generalized inverse $V^+$ contains the information of $q$.
Therefore, the ensemble mean to which all clocks synchronize explicitly depends on how the synchronization control input is fedback into controlling each clock.
This implies that the quality of the generated time scale, realized as the unobservable state, is not directly affected by the method used to estimate the observable state, or equivalently, the relative clock state.
In fact, using the Kalman filtering to estimate the observable state as in \eqref{eq:Kalobs} is not necessary to achieve the EEM synchronization.
Rather, the Kalman filtering to estimate the ``unobservable" state is important.
This will be shown later.

\subsubsection{Application to Clock Steering}\label{sec:numsteer}

The EEM synchronization algorithm in \eqref{eq:Kalent} can also be used as the steering of clocks to a reference clock.
We discuss this in more detail as follows.
Let us suppose that the $N$th clock is used as a reference clock to synchronize other clocks.
We choose the ensemble mean weight as
\[
q^{\sf T}=\mat{ 0  & \cdots & 0 &1 }.
\]
Then, the free-running ensemble mean dynamics becomes
\[
\Pi(q):
\simode{
r[k+1] & = A r[k] + v_N[k] \\
z[k] &= C r[k]
}
\]
where $v_N[k]$ denotes the process noise of the $N$th clock.
This means that the synchronization destination is set to the $N$th clock being free-running.
In fact, the control input $u_N[k]$ to the $N$th clock is always zero because
\[
u_N[k] = q^{\sf T} u[k] = q^{\sf T}V^+ \omega_o[k] =0,
\]
where the last equality is proven by \eqref{eq:kfginv}.
All other clocks are synchronized to the reference clock.

\begin{figure}[t]
\centering
\includegraphics[width = .99\linewidth]{ 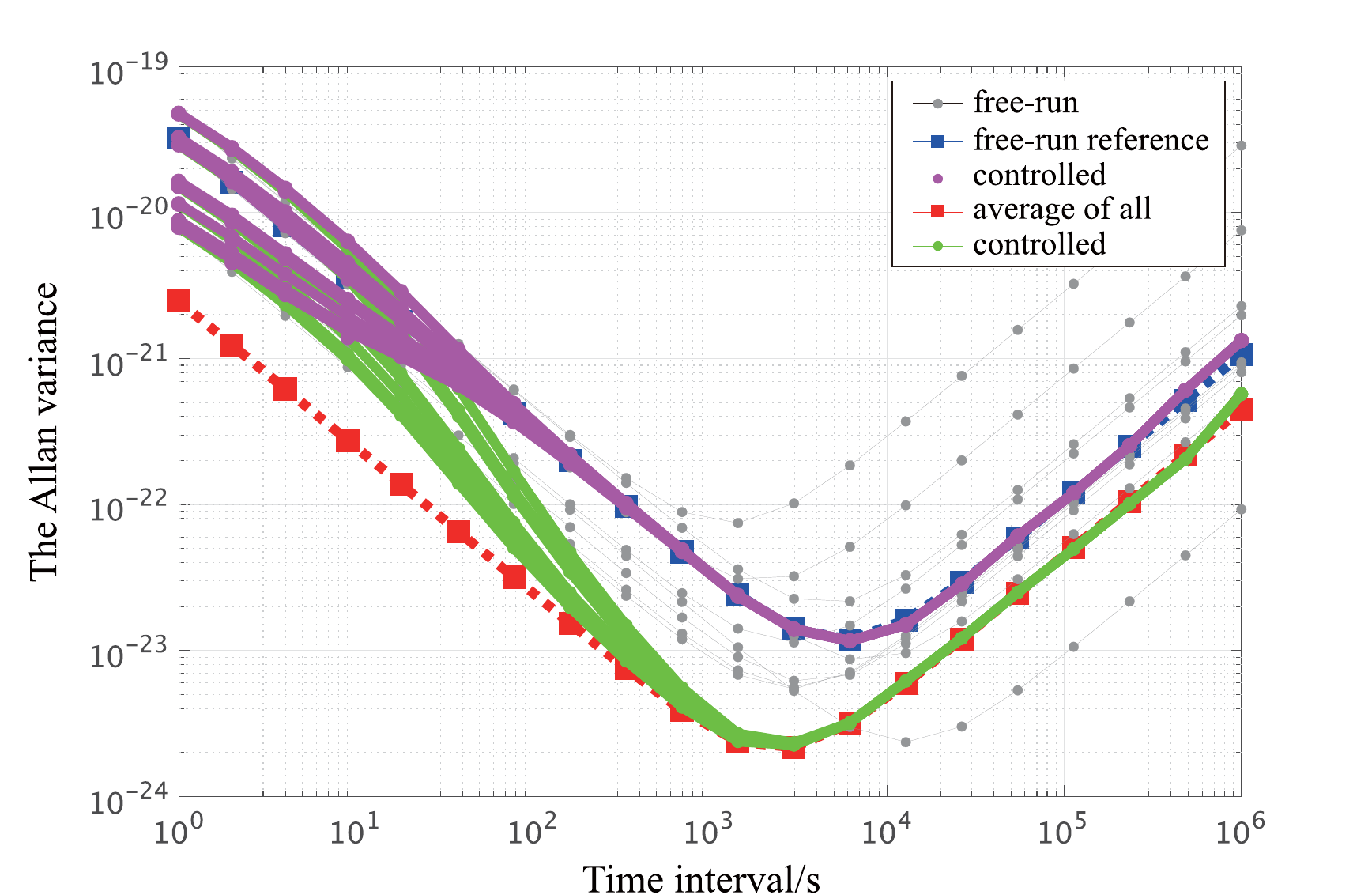}
\medskip
\caption{The Allan variances.
The grey lines are the analytical Allan variances of the free-running clocks.
The blue line is the analytical Allan variances of the free-running reference clock.
The red line is the analytical Allan variance of the average of all free-running clocks.
The purple and green lines are the statistical Allan variances of the controlled clocks.
}
\label{fig:Nthclockave}
\end{figure}

For reference, we show the result when this clock steering is applied to the ensemble clock used in Section~\ref{sec:illnum}.
We choose the feedback gain in the observable part as
\begin{equation}
\bm{F}_o = \mat{ \tfrac{0.1}{\tau} & 1 } \otimes I_9,
\end{equation}
which satisfies \eqref{eq:obsFst}.
The resulting Allan variance plots are shown in Fig.~\ref{fig:Nthclockave}.
The grey lines correspond to the free-running clocks, which are the same as those in Fig.~\ref{fig:Kalallan}.
The blue lines correspond to the free-running $N$th clock.
The purple lines correspond to the controlled clocks, i.e., the statistical Allan variance $\mathcal{L}\{h[k]\}_{k\in \mathds{T}}$ where $h[k]$ denotes the clock reading deviation of the controlled ensemble clock in \eqref{eq:Csensm}.
We can see that all other clocks are synchronized to the reference clock as expected.

We also show in Fig.~\ref{fig:Nthclockave} the result when we choose
\[
q = \frac{1}{10}\mathds{1}_{10},
\]
which means that all clocks are synchronized to the simple average.
The red lines correspond to the average of all free-running clocks.
The green lines correspond to the controlled clocks.
In the short term, we can see that the controlled clock is generally more stable than any single free-running clock because there is little variation in stability between individual clocks. 
However, in the long term, the stability of controlled clocks is lower than that of the most stable free-running clock because of the large variations in stability among individual clocks.
The next subsection will analyze the optimal ensemble mean weight that yields the least Allan variance in the short and long term.


\subsection{Least Allan Variance Synchronization Destination}

\subsubsection{Optimal Weight for Least Allan Variance}

We aim at finding an optimal ensemble mean weight that minimizes the Allan variance of the free-running ensemble mean dynamics.
We denote the Allan variance as
\begin{equation}\label{eq:allanzeta2}
\sigma_{\sf A}^2 \bigl(\tau; \Pi(q) \bigr) :=  \mathbb{E} \left[
\frac{ \left( \triangle_{1}^{2} z[k] \right)^2 }{2\tau^2}
\right],
\end{equation}
where $\Pi(q)$ in \eqref{eq:syncdes} should be considered as an implicit function of the sampling interval $\tau$.
Then, the optimal ensemble mean $q$ that gives the least Allan variance is found as follows.

\begin{lemma}\label{lem:lstAV}
Consider the free-running ensemble mean dynamics $\Pi(q)$ in \eqref{eq:syncdes}.
Define 
\begin{equation}\label{eq:Gamtau}
\varGamma(\tau) := \tau \Sigma_1 + \frac{\tau^3}{3} \Sigma_2.
\end{equation}
Then, the Allan variance in \eqref{eq:allanzeta2} is given as
\begin{equation}\label{eq:allanq}
\sigma_{\sf A}^2 \bigl(\tau; \Pi(q) \bigr) =
\frac{q^{\sf T}  \varGamma(\tau) q}{\tau^2} .
\end{equation}
Furthermore, the optimal ensemble mean weight
\[
q_{\sf A}(\tau) :=  \arg \min_{q} \sigma_{\sf A}^2 \bigl(\tau; \Pi(q) \bigr)
\quad {\rm s.t.} \quad
q^{\sf T} \mathds{1}_N =1,
\]
which minimizes the Allan variance, is given as
\begin{equation}\label{eq:optqvec2}
q_{\sf A}(\tau) = 
\frac{\varGamma^{-1}(\tau) \mathds{1}_N}{\mathds{1}_N^{\sf T} \varGamma^{-1}(\tau) \mathds{1}_N }.
\end{equation}
\end{lemma}

\begin{IEEEproof}
By standard calculation, we have
\[
\triangle_1^{2} z[k] =
\underbrace{
\mat{
C(A-2 I_2) & C 
}
}_{G}
\mat{
(I_2 \otimes q^{\sf T}) v[k] \\
(I_2 \otimes q^{\sf T}) v[k+1]
},
\]
where we have used the fact that
\[
A^2 - 2A + I_2 = 0.
\]
Therefore, the Allan variance can be written as
\[
\sigma_{\sf A}^2 \bigl(\tau; \Pi(q) \bigr)  =
\frac{1}{2\tau^2}
G
\left\{
I_2 \otimes \left[ (I_2 \otimes q^{\sf T})\bm{Q} (I_2 \otimes q) \right]
\right\}
G^{\sf T},
\]
which leads to \eqref{eq:allanq}.
Furthermore, by the Lagrange multiplier method, the ensemble mean weight $q$ that minimizes the Allan variance subject to \eqref{eq:normqkf} is found as the solution to
\[
\varGamma(\tau) q -\lambda \mathds{1}_N =0
,\quad
q^{\sf T} \mathds{1}_N =1.
\]
Thus, the minimizer is obtained as $q_{\sf A}(\tau)$ in \eqref{eq:optqvec2}.
\end{IEEEproof}
\medskip

Lemma~\ref{lem:lstAV} gives the optimal ensemble mean weight that minimizes the Allan variance of the free-running unobservable state dynamics.
In particular, to generate a time scale with best short term stability, it is reasonable to choose
\begin{equation}\label{eq:q0clock}
q_0 :=
\lim_{\tau \rightarrow 0}
q_{\sf A}(\tau) = 
\frac{ \Sigma_1^{-1} \mathds{1}_N}{\mathds{1}_N^{\sf T} \Sigma_1^{-1} \mathds{1}_N },
\end{equation}
which gives the ensemble mean weighted by the inverse ratio of the white frequency noise variances.
On the other hand, to generate a time scale with best long term stability, it is reasonable to choose
\begin{equation}\label{eq:qKclock}
q_{\infty}:= 
\lim_{\tau \rightarrow \infty}
q_{\sf A}(\tau) = \frac{\Sigma_2^{-1} \mathds{1}_N}{\mathds{1}_N^{\sf T} \Sigma_2^{-1} \mathds{1}_N },
\end{equation}
which gives the ensemble mean weighted by the inverse ratio of the random walk frequency noise variances.

\subsubsection{Need to Balance Short- and Long-Term Stability}\label{sec:needsl}

\begin{figure}[t]
\centering
\includegraphics[width = .99\linewidth]{ 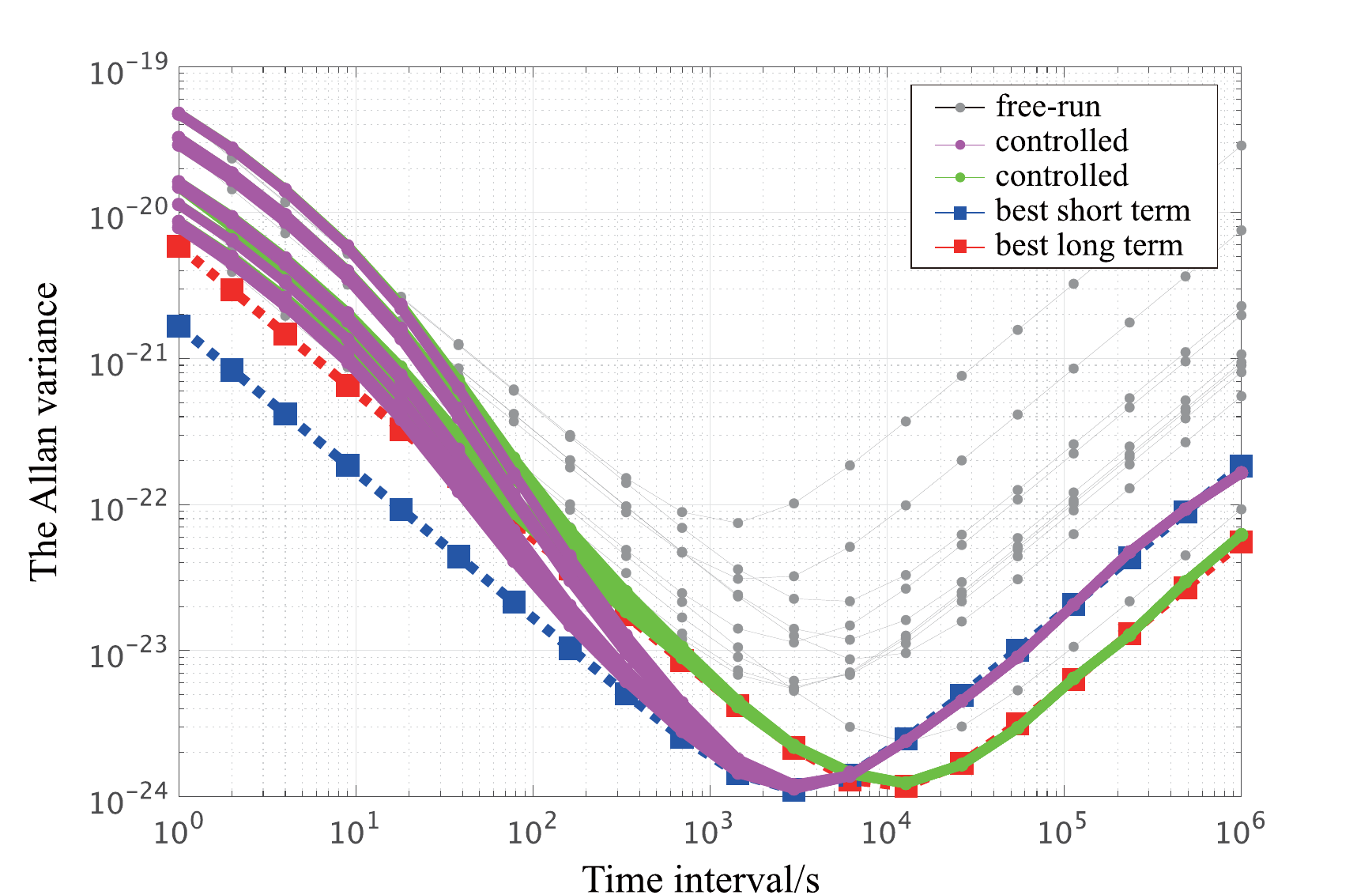}
\medskip
\caption{The Allan variances.
The grey lines are the analytical Allan variance of the free-running clocks.
The blue line is the analytical Allan variance of the best reference time scale in short term.
The red line is that in long term.
The purple and green lines are the statistical Allan variance of the controlled clocks.
}
\label{fig:clockstsl}
\end{figure}


We remark that, unless $\Sigma_1$ is a constant multiple of $\Sigma_2$, the ensemble mean weight $q_0$ for best short-term frequency stability is not equal to $q_{\infty}$ for best long-term stability.
Therefore, it is generally not possible to achieve both short- and long-term stability if the feedback gain $F_{\bar{o}}[k]$ is zero.

As an illustration, using the same ensemble clock as in Sections~\ref{sec:illnum} and \ref{sec:numsteer}, we show in Fig.~\ref{fig:clockstsl} the results when we choose $q_0$ and $q_{\infty}$.
The blue and red lines correspond to the analytical Allan variances $\sigma_{\sf A}^2 (\tau; \Pi(q_0) )$ and $\sigma_{\sf A}^2 (\tau; \Pi(q_{\infty}) )$ in \eqref{eq:allanq}, respectively.
The purple and green lines correspond to the statistical Allan variances $\mathcal{L}\{h[k]\}_{k\in \mathds{T}}$ of the controlled clocks.
This observation motivates us to achieve a better balance between short- and long-term stability by properly selecting a non-zero feedback gain $F_{\bar{o}}[k]$ for the regulation of the unobservable state dynamics.

Notably, the envelope of the blue and red lines, which represent the reference timescales with optimal short- and long-term stability, is nearly identical to the result of Kalman filtering modified by the KPW algorithm, as shown by the green line in Fig.~\ref{fig:Kalallan}.
The objective of the next subsection is to realize this time scale by regulating the frequency of the physical oscillator.

\subsection{Balancing Short and Long Term Stability}

\subsubsection{Kalman Filtering As Explicit Ensemble Mean}

To achieve a better balance of short- and long-term stability, we first prove the following important fact that the determinate Kalman filtering algorithm produces a time scale with the ``best" long-term stability.

\begin{theorem}\label{thm:kalqinf}
Consider the stationary version of the determinate Kalman filtering algorithm in \eqref{eq:Kalent}.
Then
\begin{equation}\label{eq:Hunzero}
H_{\bar{o}}^{\star}  = 0
\end{equation}
is satisfied for the unobservable state estimation if and only if the ensemble mean weight is chosen as
\begin{equation}\label{eq:qqinf}
q = q_{\infty},
\end{equation}
where $q_{\infty}$ is defined as in \eqref{eq:qKclock}.
Furthermore, the stationary error covariances of the unobservable part is given as
\begin{equation}\label{eq:Pboost}
P_{\bar{o}o}^{\star} =
\mat{
0 & - q_{\infty}^{\sf T} \Sigma_1V^{\sf T} \\
0 & 0 
}.
\end{equation}
\end{theorem}

\begin{IEEEproof}
We first prove that \eqref{eq:Hunzero} implies \eqref{eq:qqinf}.
We see from \eqref{eq:stkalgain} that, if \eqref{eq:Hunzero} holds, then
\[
P_{\bar{o}o}^{\star} \bm{C}_o^{\sf T} =0,
\]
which means that $P_{\bar{o}o}^{\star}$ is structured as
\[
P_{\bar{o}o}^{\star} =
\mat{
0 & p_1^{\sf T} \\
0 & p_2^{\sf T} 
},
\]
where $p_i$ is a vector.
Furthermore, we see from \eqref{eq:Pricc} that
\[
P_{\bar{o}o}^{\star} = Q_{\bar{o}} + AP_{\bar{o}o}^{\star} \bm{A}_o^{\sf T},
\]
which can be rewritten as
\[
\mat{
q^{\sf T}\varGamma(\tau)V^{\sf T}   & \frac{\tau^2}{2} q^{\sf T} \Sigma_2 V^{\sf T}  \\
\frac{\tau^2}{2} q^{\sf T} \Sigma_2 V^{\sf T} & \tau q^{\sf T} \Sigma_2 V^{\sf T}
}
=
\mat{
-\tau(p_1^{\sf T} + \tau p_2^{\sf T}) & - \tau p_2^{\sf T} \\
- \tau p_2^{\sf T} & 0
}.
\]
Thus, the equality of the (2,2)-block leads to \eqref{eq:qqinf}.

Conversely, if \eqref{eq:qqinf} holds, then we have
\[
Q_{\bar{o}}
=
\mat{
\tau q_{\infty}^{\sf T} \Sigma_1 V^{\sf T}   & 0  \\
0 & 0
}.
\]
By substitution, it can be verified that \eqref{eq:Pboost} is the solution to the algebraic equation
\[
P_{\bar{o}o}^{\star} = Q_{\bar o} +
A P_{\bar{o}o}^{\star}  \bigl(
I_{2(N-1)} - \bm{H}_o^{\star} \bm{C}_o 
\bigr)^{\sf T} \bm{A}_o^{\sf T} 
\]
for any $\bm{H}_o^{\star}$.
Therefore, we obtain
\[
H_{\bar{o}}^{\star}  = P_{\bar{o}o}^{\star} \bm{C}_o^{\sf T}
\bigl( \bm{C}_o \bm{P}_{oo}^{\star} \bm{C}_o^{\sf T} + \bm{R} \bigr)^{-1} =0.
\]
This proves the claim.
\end{IEEEproof}
\medskip

Theorem~\ref{thm:kalqinf} shows that the standard Kalman filtering for time scale generation can be understood as a special case of the EEM synchronization.
More specifically, if we choose the ensemble mean weight $q$ as  in \eqref{eq:qqinf}, then the synchronization control algorithm in \eqref{eq:Kalent} with $F_{\bar{o}}[k]$ equal to zero can achieve the best long-term stability.
This conversely implies that the Kalman filtering for time scale generation is generally not optimal in terms of the short-term stability, evaluated by the Allan variance.
This is a solution to {\bf Q3}.

\subsubsection{Kalman Gain for Unobservable State Estimation}

As an application of Theorem~\ref{thm:kalqinf}, we can find a simple algorithm to compute the Kalman gain $H_{\bar{o}}^{\star}$ for the unobservable state estimation in a general choice of the basis.
For illustration, we represent the basis expansion associated with the weight vector $q_{\infty}$ as
\begin{equation}\label{eq:basisqinf}
x[k] = (I_2 \otimes V^+_{\infty}) \zeta_o[k]
+ (I_2 \otimes \mathds{1}_N) \zeta_{\bar{o}}[k]
\end{equation}
where $V^+_{\infty}$ denotes the generalized inverse of $V$ associated with $q_{\infty}$.
In particular, $V^+_{\infty}$ can be defined as
\[
V^+_{\infty} := W_{\infty} (VW_{\infty})^{-1}
\]
where $W_{\infty}$ is a column full rank matrix such that 
\[
q_{\infty}^{\sf T}W_{\infty} =0.
\]
Then, we can see from \eqref{eq:kfstex} that
\begin{equation}\label{eq:basisqinfq}
\mat{
\xi_o[k] \\
\xi_{\bar{o}}[k]
}
=
\mat{
I_{2(N-1)} & 0 \\
I_2 \otimes q^{\sf T} V_{\infty}^+ & I_2
}
\mat{
\zeta_o[k] \\
\zeta_{\bar{o}}[k]
}.
\end{equation}
Considering the Kalman gains in these two bases, we have
\[
\mat{
\bm{H}_o^{\star} \\
H_{\bar{o}}^{\star}
}
=
\mat{
I_{2(N-1)} & 0 \\
I_2 \otimes q^{\sf T} V_{\infty}^+ & I_2
}
\mat{
\bm{H}_o^{\star} \\
0
}.
\]
Therefore, $H_{\bar{o}}^{\star}$ can be computed as
\[
H_{\bar{o}}^{\star} = (I_2 \otimes q^{\sf T} V_{\infty}^+) \bm{H}_o^{\star},
\]
where $\bm{H}_o^{\star}$ can be computed as the standard Kalman gain for the observable state dynamics.
In a similar way, the error covariance $P_{\bar{o}o}^{\star}$ in a general basis can be found as
\[
P_{\bar{o}o}^{\star} = (I_2 \otimes q^{\sf T} V_{\infty}^+) P_{oo}^{\star}
+
\mat{
0 & -  q_{\infty}^{\sf T} \Sigma_1 V^{\sf T} \\
0 & 0 
}.
\]

\subsubsection{Optimization of Long-Term Stability}

As shown in \eqref{eq:kalfil2}, the Kalman filtering algorithm is independent of the particular basis choice.
Thus, the estimation of the unobservable state can be used as a regulation to improve long-term stability.
In particular, using
\[
m \mathbb{Z} := \{ m k: k \in \mathbb{Z}\},
\]
we apply a collective input $\omega_{\bar{o}}[k]$ with a longer period as
\begin{subequations}\label{eq:Funobs}
\begin{equation}
F_{\bar{o}}[k]
=\left\{
\begin{array}{cl}
K_{\bar{o}}, & k\in m \mathbb{Z} \\
0, & {\rm otherwise}.
\end{array}
\right.
\end{equation}
This means that the collective control input is applied every $m$ steps.
The resulting intermitted dynamics of the unobservable state estimation in \eqref{eq:Kalunobs} is
\[
\hat{\xi}^-_o[k + m]  =  A^m \hat{\xi}^-_o[k] 
+  \eta[k] 
+ A^{m-1} B \omega_{\bar{o}}[k]
,\quad 
k\in m \mathbb{Z},
\]
where the sum of innovations is denoted as 
\[
\eta[k]:= 
\sum_{i=0}^{m-1}A^{m-1-i}H_{\bar{o}}^{\star}
\bigl( y[k+i] - \bm{C}_o \hat{\xi}^-_o[k+i]   \bigr),
\]
which is known to be a white Gaussian process; 
see \cite[Chapter~4]{aastrom2012introduction} for example.
To stabilize the unobservable state estimation dynamics, the feedback gain $K_{\bar{o}}$ is chosen such that
\begin{equation}\label{eq:unobsFst}
\bm{\Lambda} (A^{m} - A^{m-1} B K_{\bar{o}}) \subseteq \mathds{D},
\end{equation}
\end{subequations}
which is equivalent to
\[
\{ \hat{\xi}^-_{\bar{o}}[ k ] \}_{k \in m \mathbb{Z}} \in \mathcal{G}.
\]
We can see that
\[
A^m=\mat{
1 & m\tau \\
0 & 1
}
,\quad
A^{m-1}B=\mat{
 m\tau \\
 1
},
\]
which correspond to the system matrices obtained with the sampling interval $m\tau$.

Because of the intermittency of the collective control input to the unobservable state dynamics, it is clear that
\[
\xi_{\bar{o}}[k + 1]  =  A \xi_{\bar{o}}[k] + (I_2 \otimes q^{\sf T}) v[k] 
,\quad 
k\notin m \mathbb{Z}.
\]
This means that, if $q_0$ in \eqref{eq:q0clock} is chosen as the ensemble mean weight, then during the period when no collective control input is given, the unobservable state dynamics follows the dynamics of $\Pi(q_0)$, which has the best short-term stability.

On the other hand, owing to the collective control input, the unobservable state dynamics in the long term follows the dynamics of $\Pi(q_{\infty})$ with the best long-term stability.
This fact is proven as follows.

\begin{theorem}\label{thm:Kalmanq}
For the ensemble clock model in \eqref{eq:Csensm}, consider the observable canonical decomposition in \eqref{eq:qdecsys}.
Suppose that the ensemble mean weight $q$ is not equal to $q_{\infty}$ in \eqref{eq:qKclock}.
Then, the synchronization control algorithm in \eqref{eq:Kalent} achieves 
\begin{equation}\label{eq:xsyncm}
 \{ \delta [ k ] \}_{k \in m \mathbb{Z}} \in \mathcal{G}
\end{equation}
for the free-running ensemble mean dynamics $\Pi(q_{\infty})$ in \eqref{eq:syncdes} 
if and only if the feedback gains $\bm{F}_o$ and $F_{\bar{o}}[k]$ are given as in \eqref{eq:obsFst} and \eqref{eq:Funobs}, respectively.
\end{theorem}

\begin{IEEEproof}
Consider the basis expansion in \eqref{eq:basisqinf} associated with the ensemble mean weight $q_{\infty}$.
Correspondingly, we consider the expansion of the control input as
\[
u[k] = V_{\infty}^+ \nu_o[k] + \mathds{1}_N \nu_{\bar{o}}[k].
\]
Then, $\delta[k]$ in \eqref{eq:deltak} can be represented as
\[
\delta[k] = (I_2 \otimes V^+_{\infty}) \zeta_o[k]
+ (I_2 \otimes \mathds{1}_N) ( \zeta_{\bar{o}}[k] -r[k] ).
\]
Because $\zeta_{\bar{o}}[k]$ and $r[k]$ follow the dynamics of
\[
\spliteq{
\zeta_{\bar{o}}[k+1]  &= A \zeta_{\bar{o}}[k] + (I_2 \otimes q_{\infty}^{\sf T}) v[k] 
+ B \nu_{\bar{o}}[k], \\ 
r[k+1] &=  A r[k] + (I_2 \otimes q_{\infty}^{\sf T}) v[k] ,
}
\]
we can see that the difference state
\[
\hat{\zeta}_{\bar{o}}^-[k] := \zeta_{\bar{o}}[k] - r[k]
\]
follows the dynamics of  
\[
\hat{\zeta}_{\bar{o}}^-[k+1] = A \hat{\zeta}_{\bar{o}}^-[k] + B \nu_{\bar{o}}[k].
\]
This corresponds to the Kalman filtering algorithm of the unobservable state $\zeta_{\bar{o}}[k]$ because the Kalman gain of the unobservable part must be zero, as shown in Theorem~\ref{thm:kalqinf}.
Therefore, the whole Kalman filtering algorithm in this basis choice is found as
\[
\simode{
\zeta_o[k +1] &= \bm{A}_o \zeta_o[k] + (I_2 \otimes V)v[k]
+ \bm{B}_o \nu_o[k] \\
\hat{\zeta}_o^-[k+1] &= \bm{A}_o \hat{\zeta}_o^-[k]  \!+\! \bm{B}_o \nu_o[k] 
\!+\! \bm{A}_o \bm{H}_o^{\star} 
\bigl(y[k] \!-\! \bm{C}_o \hat{\zeta}_o^-[k] \bigr)\\
\hat{\zeta}_{\bar{o}}^-[k+1] &= A \hat{\zeta}_{\bar{o}}^-[k] + B \nu_{\bar{o}}[k] \\
y[k] &= \bm{C}_o \zeta_o[k] + w[k].
}
\]
Note that this system of difference equations is equivalent to the system of \eqref{eq:qdecsys} and \eqref{eq:Kalent}.
In particular, the system states and their estimates satisfy the one-to-one relation as in \eqref{eq:basisqinfq}.
In the same way, the control inputs satisfy 
\[
\mat{
\nu_o[k] \\
\nu_{\bar{o}}[k]
}
=
\mat{
I_{N-1} & 0 \\
q_{\infty}^{\sf T} V^+ & 1
}
\mat{
\omega_o[k] \\
\omega_{\bar{o}}[k]
}.
\]
Note that $q_{\infty}^{\sf T} V^+$ is not zero because $q$ is not equal to $q_{\infty}$.
Therefore, the entire system of difference equations is internally stabilized at every $m$ steps if and only if the feedback gains $\bm{F}_o$  and $F_{\bar{o}}[k]$ are given as in \eqref{eq:obsFst} and \eqref{eq:Funobs}, respectively.
This proves the claim.
\end{IEEEproof}
\medskip

Theorem~\ref{thm:Kalmanq} shows that long-term stability is optimized by feeding back the estimate of the unobservable state, regardless of the choice of the ensemble mean weight $q$.
To achieve a better balance between short- and long-term stability, we propose to choose the ensemble mean weight as $q_0$ so that the unobservable state dynamics follows $\Pi(q_0)$ in the short term, while following $\Pi(q_{\infty})$ in the long term.
This is a solution to {\bf Q4}.


\subsubsection{Demonstration of Optimal Time Scale Generation}

Using the same ensemble clock as in Sections~\ref{sec:illnum}, \ref{sec:numsteer}, and \ref{sec:needsl}, we show the result when we optimize both short- and long-term stability.
With the sampling interval of 1~[s], we consider applying the collective control input every 200~[s], i.e., we take $m$ as 200.
We choose the feedback gain for the unobservable state dynamics  as
\[
K_{\bar{o}} = \mat{
\tfrac{0.01}{m \tau} & 1
},
\]
which satisfies \eqref{eq:unobsFst}.
In general, the period of the collective control input should be selected based on the characteristics of clocks. 
More specifically, the period is selected based on the ratio of the variance of white frequency noise to the variance of white random walk frequency noise. This ratio determines the time interval at which each type of noise becomes dominant. 
In this paper, we select it considering that the dominance relationship between the noises changes after approximately 1,000 [s].

\begin{figure}[t]
\centering
\includegraphics[width = .99\linewidth]{ 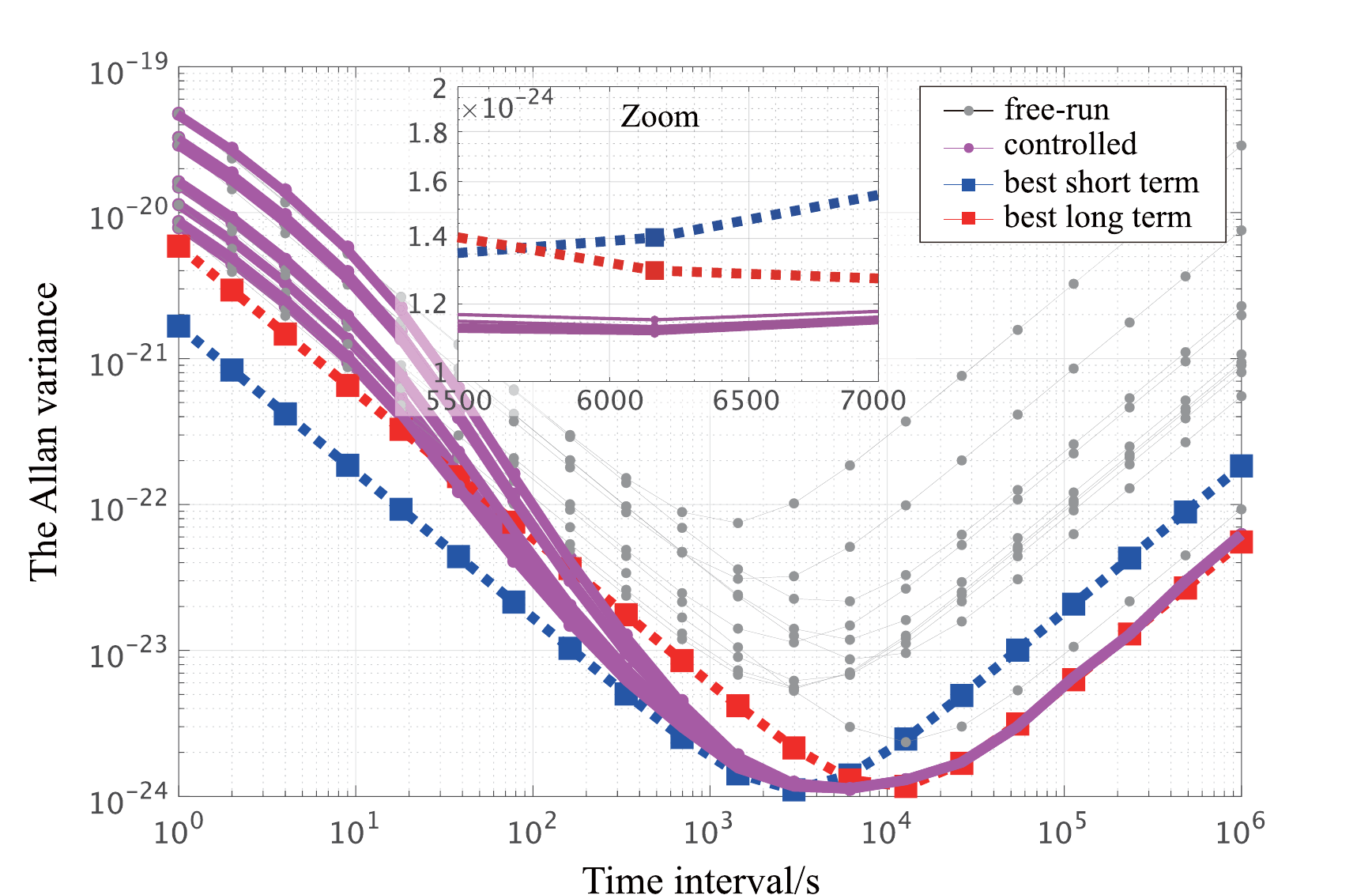}
\medskip
\caption{The Allan variances.
The grey lines are the analytical Allan variance of the free-running clocks.
The blue line is the analytical Allan variance of the best reference time scale in short term.
The red line is the analytical Allan variance of the best reference time scale in long term.
The purple lines are the statistical Allan variance of the controlled clocks.
}
\label{fig:opttsallan}
\end{figure}


The result is shown in Fig.~\ref{fig:opttsallan}.
The blue and red lines correspond to the analytical Allan variances $\sigma_{\sf A}^2 (\tau; \Pi(q_0) )$ and $\sigma_{\sf A}^2 (\tau; \Pi(q_{\infty}) )$, respectively.
The purple lines correspond to the statistical Allan variances $\mathcal{L}\{h[k]\}_{k\in \mathds{T}}$ of the controlled clocks.
Contrary to Fig.~\ref{fig:clockstsl}, the statistical Allan variances of the controlled clocks are optimized in the long term, improving both short- and long-term stability.
Notably, as shown in the zoom of Fig.~\ref{fig:opttsallan}, the Allan variance of the controlled clocks in purple shows greater stability near the time interval where the transition occurs from the short-term best reference time scale in red to the long-term best reference time scale in blue.
This suggests that achieving short-term and long-term stability yields greater stability than using constant ensemble mean weights, even with physical oscillator frequency regulation.

\begin{figure}[t]
\centering
\includegraphics[width = .99\linewidth]{ 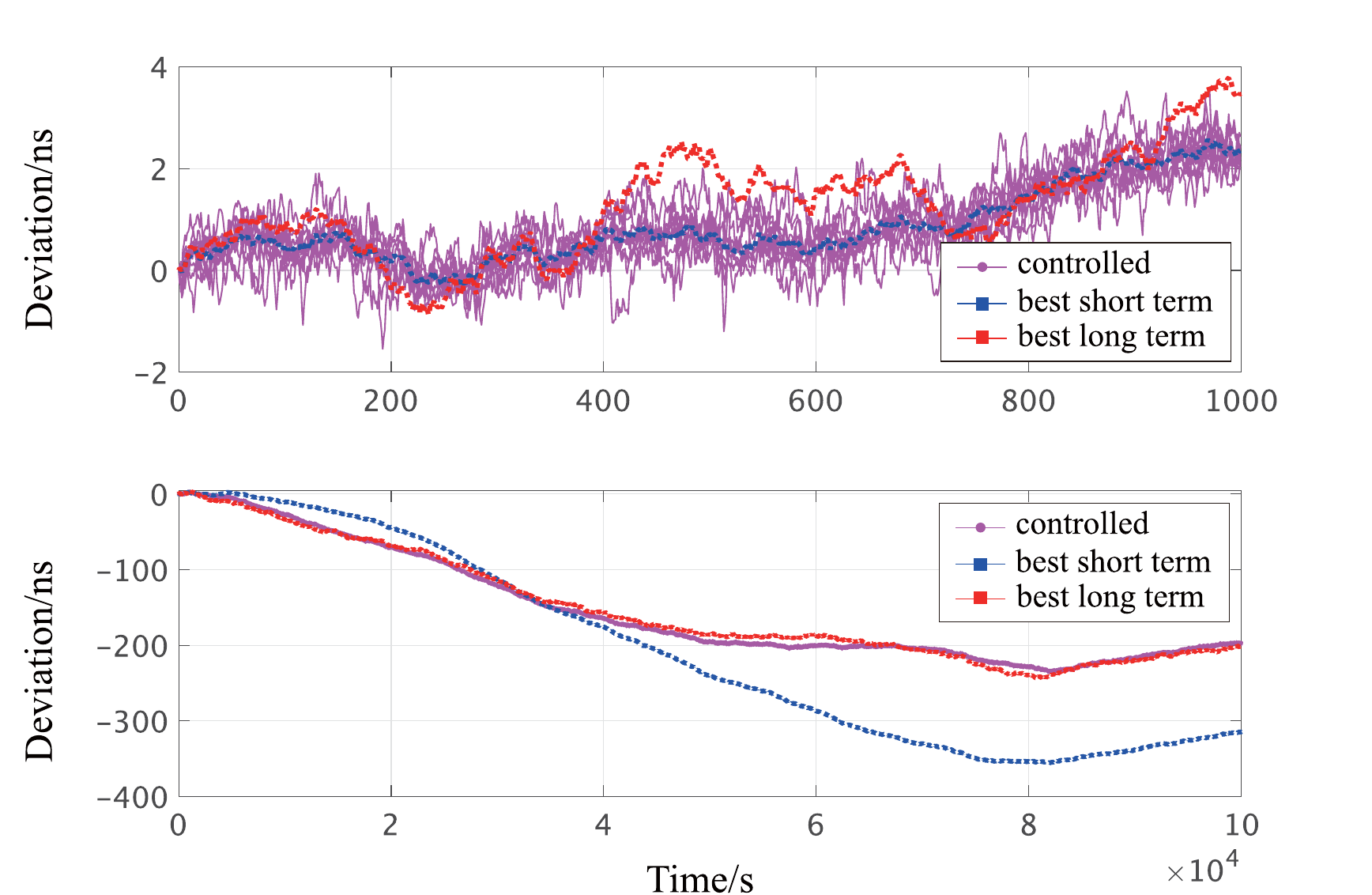}
\medskip
\caption{Time sequences of clock reading deviations.
The upper subfigure shows a short period of time, while the lower shows a long period of time.
The blue line is the best reference time scale in short term.
The red line is the best reference time scale in long term.
The purple lines are the controlled clocks.
}
\label{fig:opttimes}
\end{figure}

\begin{figure}[t]
\centering
\includegraphics[width = .99\linewidth]{ 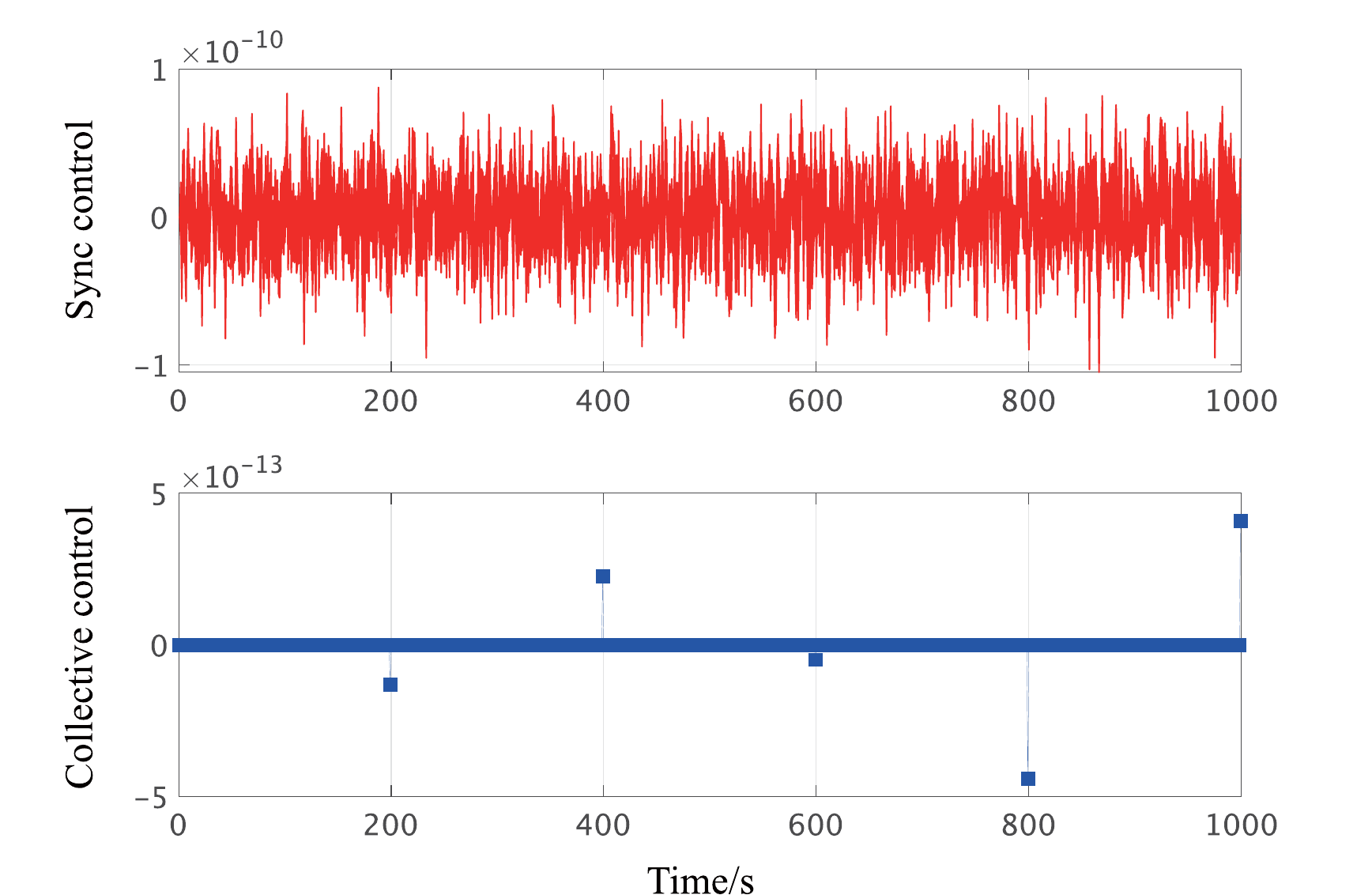}
\medskip
\caption{Input sequences.
The upper subfigure shows the control inputs for synchronization, while the lower shows the collective inputs for the regulation of synchronization destination.
}
\label{fig:optinputs}
\end{figure}

Fig.~\ref{fig:opttimes} shows the time sequences $\{h[k]\}_{k\in \mathds{T}}$ of the controlled clock reading deviations.
The upper subfigure shows the period of 1,000 [s], while the lower figure shows the period of 100,000 [s].
The blue line is the best reference time scale in short term, i.e., $\{z[k]\}_{k\in \mathds{T}}$ of $\Pi(q_0)$, while the red line is the best reference time scale in long term, i.e., $\{z[k]\}_{k\in \mathds{T}}$ of $\Pi(q_{\infty})$.
The purple lines are the controlled clocks.
We can see that the controlled clocks follow the dynamics of $\Pi(q_0)$ in the short term, while they follow the dynamics of $\Pi(q_{\infty})$ in the long term, as expected. 

Finally, we show the resulting control inputs in Fig.~\ref{fig:optinputs}.
The upper subfigure shows the synchronization control input $\{\omega_{o}[k]\}_{k\in \mathds{T}}$, while the lower shows the collective control input $\{\omega_{\bar{o}}[k]\}_{k\in \mathds{T}}$ for regulating synchronization destination.
By design, the latter is given every 200 [s].


\section{Concluding Remarks}

This paper has presented the explicit ensemble mean (EEM) synchronization framework that unifies time scale generation, clock synchronization, and oscillator frequency regulation within the systems and control theory paradigm.
Based on the observable canonical decomposition of a standard ensemble clock model, we have clarified that the unobservable state estimation by Kalman filtering plays an important role in generating the optimal time scale in the long term, as measured by the Allan variance.
The proposed framework offers a systematic approach to achieving short- and long-term stability in the synchronization destination of clocks or the generated time scale.

Though the analytical results in this paper are validated by numerical simulations under ideal Gaussian noise assumptions, our experiments using real atomic clock data and hardware implementations demonstrate the favorable performance of the proposed method, even when unmodeled dynamics may be present. 
We will report these practical results in a separate publication.

There are several important areas for future work. 
For practical operation, there must be robust handling of abnormal clocks, including detection and exclusion mechanisms, as well as the ability to adapt to changes in the ensemble. 
The proposed framework will be expanded to address these conditions.

Another important direction is to conduct a comprehensive structural analysis of existing time scale generation algorithms, such as AT1, KAS-2, and KPW, within the proposed framework. 
Including third-order clock dynamics with frequency drift and heterogeneous model orders in the ensemble will further expand the applicability. 
Additionally, integrating digital clock reading adjustment with physical oscillator frequency regulation could result in a more flexible, comprehensive timing infrastructure.

%

\section*{Acknowledgment}

This work is supported by research and development conducted by the Ministry of Internal Affairs and Communications (MIC) under its "Research and Development for Expansion of Radio Resources (JPJ000254)" program.

\ifCLASSOPTIONcaptionsoff
  \newpage
\fi

\bibliographystyle{unsrt}
\bibliography{refs}

\end{document}